\documentclass[aps,pra,reprint,superscriptaddress]{revtex4-2}

\usepackage{color}

\usepackage{amsmath}
\usepackage{bm}
\usepackage[dvipdfmx]{graphicx}

\usepackage{comment}

\newcommand{\adag}{\hat{a}^\dagger}
\newcommand{\adagtwo}{\hat{a}^{\dagger 2}}
\newcommand{\amdag}{\hat{a}^\dagger_-}
\newcommand{\amdagtwo}{\hat{a}^{\dagger 2}_-}
\newcommand{\apdag}{\hat{a}^\dagger_+}
\newcommand{\apdagtwo}{\hat{a}^{\dagger 2}_+}
\newcommand{\ahat}{\hat{a}}
\newcommand{\ahattwo}{\hat{a}^{2}}
\newcommand{\amhat}{\hat{a}_-}
\newcommand{\amhattwo}{\hat{a}^{2}_-}
\newcommand{\aphat}{\hat{a}_+}
\newcommand{\aphattwo}{\hat{a}^{2}_+}

\newcommand{\bra}[1]{\langle #1 |}
\newcommand{\Cmat}{C_{\text{mat}}}
\newcommand{\Cmatinv}[1]{\left( C_{\text{mat}}^{-1} \right)_{#1}}
\newcommand{\deltaEJ}{\delta E_J(\Phi_e, \delta \Phi)}
\newcommand{\ECgp}{E'_{C_g}}
\newcommand{\eiThetakt}{e^{i\Theta_k(t)}}
\newcommand{\EJg}{E_{J_g}}
\newcommand{\EJgtwo}{E_{J_g}^{(2)}}
\newcommand{\EJgfour}{E_{J_g}^{(4)}}
\newcommand{\EJSigma}{E_{J_\Sigma} (\Phi_e)}
\newcommand{\gfour}{g^{(4)}}
\newcommand{\gtwo}{g^{(2)}}
\newcommand{\hc}{\text{h.c.}}
\newcommand{\Hhat}{\hat{\mathcal{H}}}
\newcommand{\Km}{K_-}
\newcommand{\Kp}{K_+}
\newcommand{\ket}[1]{|#1\rangle}
\newcommand{\omegap}[1]{\omega_{p, #1}}
\newcommand{\Phidc}{\Phi_\text{dc}}
\newcommand{\phigm}{\phi_{g_-}}

\newcommand{\sumkfour}{\sum_{k=1}^{4}}
\newcommand{\sumlfour}{\sum_{l=1}^{4}}
\newcommand{\Thetakpmt}{\Theta_k^\pm(t)}
\newcommand{\Thetakpt}{\Theta_k^+(t)}
\newcommand{\Thetakmt}{\Theta_k^-(t)}
\newcommand{\Utrans}[1]{\hat{U}_r^\dagger \hat{U}_g^\dagger #1 \hat{U}_g \hat{U}_r}

\begin{document}


\title{Four-body coupler for superconducting qubits based on Josephson parametric oscillators}


\author{Ryoji Miyazaki}
\affiliation{Secure System Platform Research Laboratories, NEC Corporation, Kawasaki 211-8666}
\affiliation{NEC-AIST Quantum Technology Cooperative Research Laboratory, National Institute of Advanced Industrial Science and Technology
(AIST), Tsukuba, Ibaraki 305-8568, Japan}

\author{Tsuyoshi Yamamoto}
\affiliation{Secure System Platform Research Laboratories, NEC Corporation, Kawasaki 211-8666}
\affiliation{NEC-AIST Quantum Technology Cooperative Research Laboratory, National Institute of Advanced Industrial Science and Technology
(AIST), Tsukuba, Ibaraki 305-8568, Japan}


\date{\today}

\begin{abstract}
We theoretically propose a circuit of the four-body coupler for superconducting qubits based on Josephson parametric oscillators (JPOs). Our coupler for the four-body interaction has a superconducting loop, similar to a capacitively shunted flux qubit, where an external magnetic flux set to half a flux quantum is threaded. This coupler circuit is a specific setup of the circuit called superconducting nonlinear asymmetric inductive elements (SNAIL) and also is a generalization of the previously proposed one for the four-body interaction of JPOs. We clarify roles of circuit parameters in the four-body interaction and, in particular, show that the four-body coupling constant in our circuit can be significantly increased by tuning capacitance of the coupler or the area ratio of the Josephson junctions of the coupler.
\end{abstract}

\maketitle

\section{Introduction}

Technologies of quantum computations have significantly advanced lately, leading to demonstration of their computational advantage~\cite{F.Arute2019, P.Jurcevic2021, Y.Wu2021}. 
One factor essential to such advancements is continual improvement of couplers of qubits. The couplers are usually considered for two qubits, 
because the set of a single qubit and coupled two qubits can realize universal quantum computation~\cite{M.Nielsen2010}. 
Even in quantum annealing~\cite{T.Kadowaki1998, P.Hauke2020}, networks with two-qubit couplings are sufficient to represent any combinatorial optimization problems to be solved~\cite{F.Barahona1982}. This does not mean multi-body interactions of qubits such as three-body ones are not worth studying. They are useful to reduce the overhead in implementing some quantum algorithms and simulating many-body quantum physics~\cite{X.Wen1989, A.Kitaev2003, M.Muller2011, J.Seeley2012, R.Babbush2018}. In other words, the multi-body interactions can partially avoid the difficulty in integrating many qubits or coupling distant qubits. Therefore, the implementation of multi-body interactions can be a promising option to efficiently use limited scale quantum computers.

One application of multi-body interactions of qubits is the scheme based on parity encoding~\cite{N.Sourlas2005}, which is also called the Lechner-Hauke-Zoller (LHZ) scheme~\cite{W.Lechner2015, K.Ender2023, M.Drieb-schon2023, M.Fellner2023, R.Hoeven2024}, proposed for quantum annealing. This scheme has been intensively studied for not only quantum annealing~\cite{W.Lechner2015, M.Leib2016, F.Pastawski2016, T.Albash2016, A.Rocchetto2016, A.Glaetzle2017, N.Chancellor2017, S.Puri2017Jun, L.Sieberer2018, P.Zhao2018, C.Dlaska2019, A.Hartmann2019Apr, A.Hartmann2019Sep, Y.Susa2020, T.Kanao2021, Y.Susa2021, M.Lanthaler2021, M.Konz2021, T.Yamaji2022Jun, M.Lanthaler2023, K.Ender2023, M.Drieb-schon2023, M.Fellner2023, G.Mbeng2022arXiv, R.Hoeven2024, Y.Nambu2024FebarXiv, Y.Nambu2024JularXiv} but also gate-based quantum computation~\cite{W.Lechner2020, C.Dlaska2022, K.Ender2022, M.Fellner2022OctPRL, M.Fellner2022OctPRA, A.Weidinger2023, P.Sriluckshmy2023, J.Unger2022arXiv, R.Hoeven2023}. Its classical counterpart has also been examined~\cite{Y.Nambu2022, S.Razmkhah2024}. An attractive advantage of this scheme is to embed the all-to-all two-body interactions of logical qubits in a planar network of physical qubits only with the nearest-neighbor interactions. The required interactions of physical qubits, however, are three- or four-body ones. The implementation of such multi-body interactions is the biggest challenge to realize this scheme.

Theoretical ideas to implement the LHZ scheme have been proposed for superconducting circuits~\cite{M.Leib2016, S.Puri2017Jun, N.Chancellor2017, P.Zhao2018} and Rydberg atoms~\cite{A.Glaetzle2017, C.Dlaska2022}. One of the ideas shows the superconducting-circuit structure for the four-body interaction of qubits based on Josephson parametric oscillators (JPOs) via a coupler and demonstrates numerical simulations of quantum annealing under the LHZ scheme in the circuit~\cite{S.Puri2017Jun}. A JPO with a low photon loss rate works as a Kerr-nonlinear parametric oscillator~\cite{Z.Wang2019, A.Grimm2020, T.Yamaji2022Feb}, which generates a cat qubit~\cite{H.Goto2016Feb, H.Goto2016May, S.Puri2017Apr, A.Grimm2020, S.Masuda2021}. Thus, this idea~\cite{S.Puri2017Jun} can also be regarded as a proposition of a method to implement a four-body interaction of cat qubits. This point is noteworthy, because the cat qubits have attracted much interest, in particular, for their features like the robustness against single-photon loss and phase-flip errors~\cite{S.Puri2017Apr, A.Grimm2020, S.Puri2020, A.Darmawan2021, Q.Xu2022, H.Putterman2022, T.Kanao2022}.

The LHZ scheme requires a large coupling constant of a four-body interaction to map states of physical qubits to those of logical ones without inconsistency. We therefore need to know how we can increase the four-body coupling constant in the circuit. To this end, we should revisit the above proposition to elaborate on the circuit-parameter dependence of the coupling constant. Obtaining a detailed expression of the coupling constant will enable us to find parameter settings to increase the four-body interaction. Furthermore, it would be valuable to explore extensions of the circuit structure to achieve larger coupling constants.

In this paper, we theoretically propose a coupler for the four-body interaction of JPOs. The coupler is a generalization of the previous one~\cite{S.Puri2017Jun} by adding extra control knobs. We aim to find a coupler 
that can increase the four-body coupling constant and to clarify how to increase it on the basis of its circuit-parameter dependence and underlying physics. The next section shows our circuits with the proposed coupler. 
In Sec.~\ref{sec:derivation}, we derive the effective Hamiltonian for four-body interaction of JPOs in our circuits. The circuit-parameter dependence of its coupling constant is investigated in Sec.~\ref{sec:parameter-dependence}. The analysis in this section demonstrates how our additional control knobs increase the four-body interaction. 
Section~\ref{sec:conclusion} presents our conclusion.

\section{proposed circuit}
\label{sec:setting}

\begin{figure}
\includegraphics[width=\columnwidth]{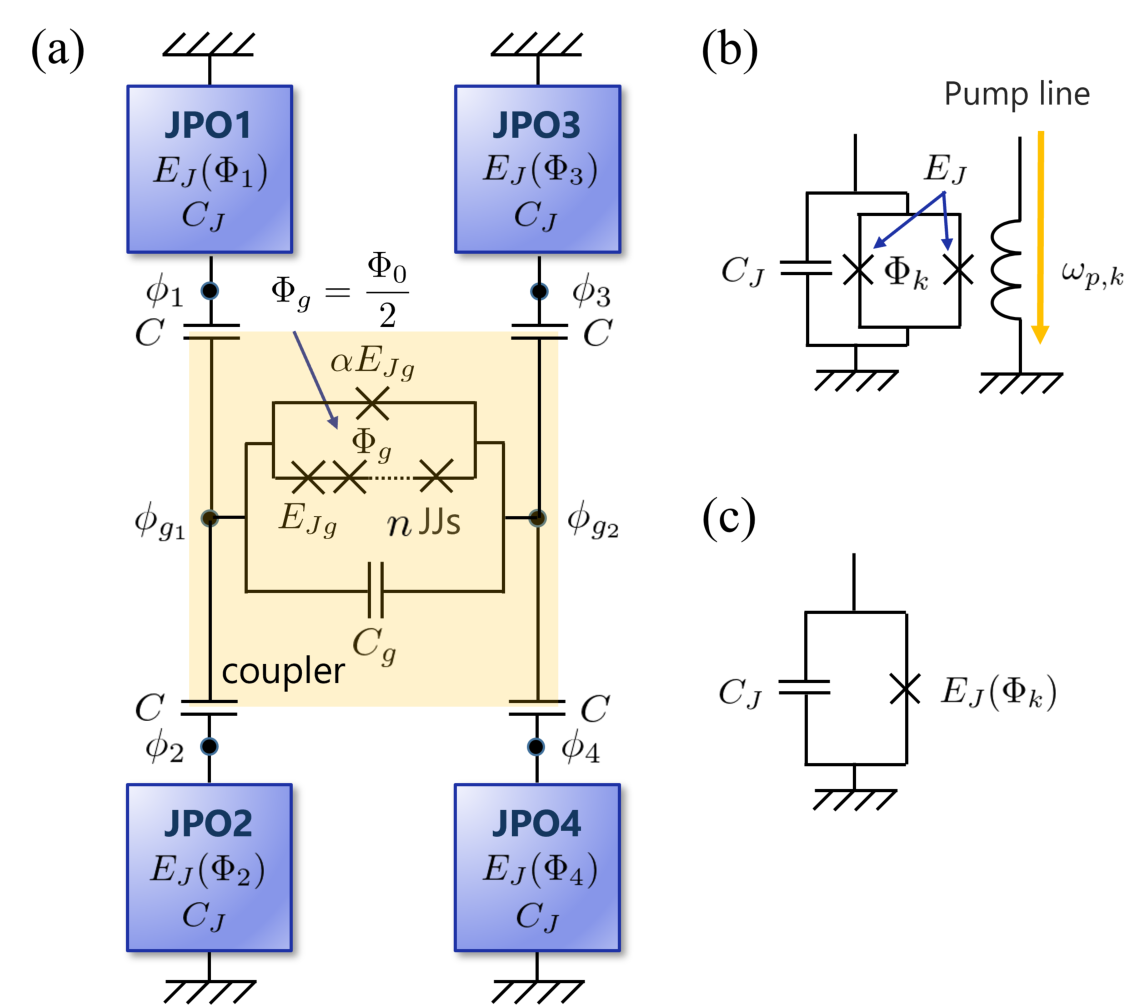}%
\caption{\label{fig:circuit}
(a) Schematic of our proposed circuit for the four-body interaction of JPOs. The circuit has four JPOs (blue squares) and a coupler (shaded part). JPOs 1 and 2 are connected to one of the two edges of the coupler with capacitance $C$, while JPOs 3 and 4 are similarly connected to the other edge. The coupler is a superconducting loop shunted by capacitance $C_g$. One branch of the loop has $n$ identical JJs with Josephson energy $E_g$ in series. 
The other has a single JJ with smaller Josephson energy $\alpha \EJg$ where $\alpha < 1$. 
We call the two branches array branch and smaller JJ branch, respectively.
The loop is threaded by an external magnetic flux $\Phi_g$ set to half a flux quantum $\Phi_0/2$. Dots indicate nodes for the node flux vector $\bm{\phi}$ in Eq.~(\ref{eq:L}). (b) JPO $k$ in our circuit. 
The current with frequency $\omegap{k}$ in the pump line induces a modulated magnetic flux $\Phi_k$ threading the SQUID of the JPO. 
The circuit is equivalent to the one shown in (c), where a JJ with Josephson energy $E_J(\Phi_k)$ [Eq.~(\ref{eq:EJPhik})] is shunted by capacitance $C_J$.
			}
\end{figure}

We first describe the superconducting circuit we propose, which has four JPOs and a coupler as depicted in Fig.~\ref{fig:circuit}(a). 
Each JPO consists of a $C$-shunted superconducting quantum interference device (SQUID) threaded by a modulated magnetic flux [Fig.~\ref{fig:circuit}(b)]. For simplicity, we assume that the four JPOs have the same capacitance $C_J$ and Josephson energy $E_J$ of each Josephson junction (JJ). 
This assumption leads to the same resonance frequency $\omega$ of the four JPOs, 
but it is not essential for deriving the four-body interaction. 
The circuit structure of a JPO can be generalized, for example, by adding JJs or linear inductors in series with the SQUID~\cite{A.Yamaguchi2024}. 
The magnetic flux $\Phi_k$ threading the SQUID of JPO $k$ is modulated as
\begin{equation}
\Phi_k 
= \Phidc + \delta \Phi \cos \left( \omega_{p,k} t + \theta_{p,k} \right), 
\label{eq:Phik}
\end{equation}
where $\Phidc$ is the static offset.
We can well approximate the effective Josephson energy of the SQUID [Fig.~\ref{fig:circuit}(c)] as
\begin{equation}
E_{J} (\Phi_k)
= \EJSigma+ \deltaEJ \cos (\omegap{k} t + \theta_{p,k}), 
\label{eq:EJPhik}
\end{equation}
where 
\begin{equation}
\EJSigma 
= 2E_J \cos \left( \pi \frac{\Phidc}{\Phi_0} \right), 
\label{eq:EJSigma}
\end{equation}
\begin{equation}
\deltaEJ 
= - 2E_J \sin \left( \pi \frac{\Phidc}{\Phi_0} \right) 
\pi \frac{\delta\Phi}{\Phi_0}.
\label{eq:deltaEJ} 
\end{equation}
We assume $|\deltaEJ| \ll \EJSigma$.
The details of Eq.~(\ref{eq:EJPhik}) are shown in Appendix~\ref{sec:effective_Josephson_energy}.
Pump frequencies $\omegap{k}$ are set to nearly double the resonance frequency of JPO $k$, that is, $\omegap{k} \simeq 2 \omega$, and satisfy 
\begin{equation}
\omegap{1} + \omegap{2} = \omegap{3} + \omegap{4}.
\label{eq:pump_frequency}
\end{equation}
We will find that this condition makes four-body coupling terms remain under the rotating-wave approximation~\cite{S.Puri2017Jun}. $\theta_{p,k}$ in Eq.~(\ref{eq:EJPhik}) will be utilized to tune the magnitude and sign of the four-body coupling term. JPOs 1 and 2 are connected to one of the two edges of the coupler with capacitance $C$, while JPOs 3 and 4 are similarly connected to the other edge. 
We assume 
$C \ll C_J$, 
where linear two-body interactions of JPOs are suppressed as discussed in Appendix~\ref{sec:classical_Hamiltonian}. 
The coupler is a superconducting loop shunted by capacitance $C_g$. 
One branch of the loop has $n$ identical JJs with Josephson energy $\EJg$ in series. 
The other has a single JJ with smaller Josephson energy $\alpha \EJg$ where $\alpha <1$. 
We call the two branches array branch and smaller JJ branch, respectively.
Each JJ is assumed to satisfy the condition that it has a charging energy much smaller than $\EJg$ and capacitance across it much larger than that to the ground.
This condition allows us to treat each phase without quantum fluctuations~\cite{K.Matveev2002, U.Vool2017}. 
An external magnetic flux $\Phi_g$ set to half a flux quantum $\Phi_0/2$ threads the loop. This magnetic flux is applied to shift the Josephson phase across the coupler by $\pi$. This shift can also be realized without the magnetic flux by using ferromagnetic Josephson $\pi$ junctions~\cite{T.Yamashita2020, S.Kim2024}.

Our circuit is a generalization of the previously proposed one~\cite{S.Puri2017Jun}. When $n = 1$ and $\alpha = 0$, the circuit reduces to the one in Ref.~\cite{S.Puri2017Jun}, except our coupler is capacitively shunted. When $\alpha = 1/n$, the coupler circuit is the same as that of the quarton~\cite{F.Yan2020arxiv}, which is one specific type of the circuit called superconducting nonlinear asymmetric inductive elements (SNAIL)~\cite{A.Zorin2016, N.Frattini2017}. The quarton is characterized by its quartic potential function without a quadratic term and has been proposed as a nonlinear coupler between two qubits/photons~\cite{Y.Ye2021, Y.Ye2024}. Its quartic potential yields a purely nonlinear coupling. Further, removing the quadratic term, which effectively reduces the coupling strength, enables stronger nonlinear coupling than that via a $C$-shunted SQUID. We utilize this structure as a four-body coupler for JPOs. Note that unlike the nonlinear coupler of two qubits/photons in Ref.~\cite{Y.Ye2021}, where the quarton is galvanically connected to the qubits, our circuit has the modes of the coupler due to its capacitances to each JPO. The JPOs interact via one of the modes of the coupler.

\section{Derivation of effective Hamiltonian}
\label{sec:derivation}

The method of nodes~\cite{U.Vool2017} gives the Lagrangian of the circuit shown in Fig.~\ref{fig:circuit}(a) as
\begin{equation}
\mathcal{L} 
=\frac{1}{2} \dot{\bm{\phi}}^T \Cmat \dot{\bm{\phi}}
- V(\bm{\phi}), 
\label{eq:L}
\end{equation}
where $\bm{\phi} = (\phi_1, \phi_2, \phi_3, \phi_4, \phi_{g_1}, \phi_{g_2})^T$ is the node flux vector, 
and $\Cmat$ is the capacitance matrix whose elements are described in Appendix~\ref{sec:classical_Hamiltonian}. $V(\bm{\phi})$ is the potential function, which is the sum of those for JPOs and the coupler,
\begin{equation}
V(\bm{\phi})
= \sumkfour V_k (\phi_k) + V_g (\phi_{g_1}, \phi_{g_2}).
\end{equation}
For the JPOs, we have
\begin{equation}
V_k (\phi_k) 
= - E_J(\Phi_k) \cos \frac{\phi_k}{\phi_0},
\end{equation}
where $E_J(\Phi_k)$ is the effective Josephson energy given by Eq.~(\ref{eq:EJPhik}), and $\phi_0 = \hbar / 2e$. 
The coupler has a superconducting loop 
cosisting of $n$ identical JJs with Josephson energy $\EJg$ in series 
and a single JJ with smaller Josephson energy $\alpha \EJg$.
Thus, its potential function is
\begin{eqnarray}
&&V_g (\phi_{g_1}, \phi_{g_2})
\nonumber
\\
&&= - n\EJg \cos \frac{\phigm}{n\phi_0}
- \alpha \EJg \cos \left( \frac{\phigm}{\phi_0} + 2\pi \frac{\Phi_g}{\Phi_0} \right),
\end{eqnarray}
where we have defined $\phi_{g_{\pm}} = \phi_{g_1} \pm \phi_{g_2}$.
Since the external magnetic flux $\Phi_g$ is set to half a flux quantum $\Phi_0/2$, 
the potential function is rewritten as
\begin{eqnarray}
V_g(\phi_{g_1}, \phi_{g_2})
&&= - n\EJg \cos \frac{\phigm}{n\phi_0}
+ \alpha \EJg \cos \frac{\phigm}{\phi_0}
\nonumber
\\
&&= \sum_{m=0}^\infty
\frac{(-1)^{m+1} \EJg^{(2m)}}{(2m)!}\left(\frac{\phigm}{\phi_0}\right)^{2m}, 
\label{eq:Ug}
\end{eqnarray}
where
\begin{equation}
\EJg^{(m)} = \left( \frac{1}{n^{m-1}} - \alpha \right) \EJg.
\label{eq:EJgm}
\end{equation}
Note that $\EJg^{(m)}$ does not actually depend on $m$ when $n = 1$.

The Legendre transformation gives the corresponding Hamiltonian, 
\begin{equation}
\mathcal{H}
= \frac{1}{2} \bm{q}^T C_{\text{mat}}^{-1} \bm{q} + V(\bm{\phi}), 
\label{eq:H_formal}
\end{equation}
where $\bm{q} = (q_1, q_2, q_3, q_4, q_{g_1}, q_{g_2})^T = C_{\text{mat}}\dot{\bm{\phi}}$. The elements of $C_{\text{mat}}^{-1}$ are shown in Appendix~\ref{sec:classical_Hamiltonian}. We assume 
that the coupling capacitance $C$ between a JPO and the coupler is much smaller 
than the capacitance $C_J$ sunting the SQUID of a JPO, 
$C \ll C_J$, 
as mentioned in Sec.~\ref{sec:setting} 
and therefore ignore terms of $O(C/C_J)$. 
The Hamiltonian reads 
\begin{eqnarray}
\mathcal{H}
=&& \frac{1}{2C_J} \sumkfour q_k^2
\nonumber
\\
&&+ \frac{1}{2} \left( \frac{1}{C} + \frac{1}{C_J} \right) q_{g_+}^2 
+ \frac{1}{2C_J} \left[ \frac{C_J}{C_g + C} + \left( \frac{C}{C_g + C} \right)^2 \right] q_{g_-}^2
\nonumber
\\
&&+ \frac{1}{2C_J} \sumkfour q_k q_{g_+}
+ \frac{C}{2C_J(C_g + C)} \sumkfour s_k q_k q_{g_-}
\nonumber
\\
&&+ V(\bm{\phi}),
\label{eq:H} 
\end{eqnarray}
where $q_{g_\pm} = (q_{g_1} \pm q_{g_2} )/2$ and 
\begin{equation}
s_1 = s_2 = 1, 
\ \ \ 
s_3 = s_4 = -1.
\label{eq:s}
\end{equation}
The detailed calculations are shown in Appendix~\ref{sec:classical_Hamiltonian}.

To derive the Hamiltonian for the corresponding quantum system, 
we introduce operators for the phase $\phi_k/\phi_0$ 
and the number $q_k/2e$ of Cooper pairs for $k = 1, 2, 3, 4, g_+$, and $g_-$ 
as 
\begin{equation}
\hat{\varphi}_k = \varphi_{kZ} \left( \ahat_k + \adag_k \right),
\label{eq:varphik}
\end{equation}
\begin{equation}
\hat{n}_k = -i n_{kZ} \left( \ahat_k - \adag_k \right),
\end{equation}
respectively, where 
\begin{equation}
\varphi_{kZ} = \sqrt{\frac{Z_k}{2}},
\label{eq:varphikZ} 
\end{equation}
\begin{equation}
n_{kZ} = \sqrt{\frac{1}{2Z_k}}. 
\end{equation}
$\adag_k$ and $\ahat_k$ are bosonic creation and annihilation operators, respectively, 
leading to $[\hat{\varphi}_k, \hat{n}_k] = i$. 
$Z_k$ is defined by 
\begin{equation}
Z_k = \sqrt{\frac{8E_C}{\EJSigma}} 
\end{equation}
for $k = 1, 2, 3$, and 4, 
where 
\begin{equation}
E_C = \frac{e^2}{2C_J},
\label{eq:EC} 
\end{equation}
and 
$\EJSigma$ is the static Josephson energy of a JPO 
[Eq.~(\ref{eq:EJSigma})].
We also have
\begin{equation}
Z_{g_+} = Z_{g_-} = \sqrt{\frac{8\ECgp}{\EJgtwo}}, 
\label{eq:Zgpm}
\end{equation}
where
\begin{equation}
\ECgp
= \frac{e^2}{2C_J} \left[ \frac{C_J}{C_g + C} + \left( \frac{C}{C_g + C} \right)^2 \right].
\label{eq:ECg'}
\end{equation}
We replace $\varphi_k$ and $q_k/2e$ in $\mathcal{H}$ [Eq.~(\ref{eq:H})] with $\hat{\varphi}_k$ and $\hat{n}_k$, respectively, to obtain
\begin{widetext}
\begin{eqnarray}
\frac{\Hhat}{\hbar}
=&& \sumkfour 
\left[ 
\omega \adag_k \ahat_k 
- \frac{K}{12} \left( \ahat_k + \adag_k \right)^4 
+ p \left( \ahat_k + \adag_k \right)^2 \cos \left( \omegap{k} t + \theta_{p,k} \right)
\right]
\nonumber
\\
&&+ \omega_+ \apdag \aphat 
- \frac{\omega_+}{2} \left( \aphat^2 + \apdagtwo \right)
+ \omega_- \amdag \amhat 
- \frac{\Km}{12} \left( \amhat + \amdag \right)^4
\nonumber
\\
&&+ g_+ \sumkfour \left( \adag_k - \ahat_k \right) \left( \aphat - \apdag \right) 
+ g_- \sumkfour s_k \left( \adag_k - \ahat_k \right) \left( \amhat - \amdag \right), 
\label{eq:Hhat}
\end{eqnarray}
\end{widetext}
where
\begin{equation}
\omega
= \frac{\sqrt{8 E_C \EJSigma}}{\hbar},
\label{eq:omega}
\end{equation}
\begin{equation}
K
= \frac{E_C}{\hbar},
\label{eq:K} 
\end{equation}
\begin{equation}
p
= \frac{\deltaEJ \omega}{4\EJSigma}
\end{equation}
\begin{equation}
\omega_+ 
= 4\frac{E_C}{\hbar} \left(\frac{C_J}{C} + 1 \right) \sqrt{\frac{\EJgtwo}{8\ECgp}}, 
\end{equation}
\begin{equation}
\omega_-
= \frac{\sqrt{8 \ECgp \EJgtwo}}{\hbar}, 
\label{eq:omega-}
\end{equation}
\begin{equation}
\Km
= \frac{\ECgp \EJgfour}{\hbar \EJgtwo}, 
\label{eq:Kg-}
\end{equation}
\begin{equation}
g_+ 
= \frac{\sqrt{\omega \omega_-}}{4} \sqrt{\frac{E_C}{\ECgp}}, 
\end{equation}
and
\begin{equation}
g_- 
= \frac{C}{C_g + C} g_+.
\label{eq:g-}
\end{equation}
We have taken into account up to the fourth order of $\varphi_{kZ}$ but ignored constants and $\deltaEJ/\EJSigma \varphi_{kZ}^4$, since we set $|\deltaEJ| \ll \EJSigma$.
$\omega$, $K$, and $p$ are the resonance frequency, the Kerr coefficient, 
and the pump amplitude, respectively, of a JPO.
These parameters are common to all JPOs in our simple setting.
$\omega_+$ and $\omega_-$ are the resonance freqencies 
for coupler's two modes $\aphat$ and $\amhat$.
These two modes couple to the modes of each JPO with the coupling constants $g_+$ and $g_-$, respectively.

We next perturbatively incorporate effects of the coupler into the JPOs to derive the effective Hamiltonian that describes the direct coupling between the JPOs. To this end, we transform $\Hhat$ with a unitary operator
\begin{equation}
\hat{U}_g
= e^{\hat{S}}, 
\label{eq:U_g}
\end{equation}
\begin{eqnarray}
\hat{S} 
= - g'_+  && \left[ \sumkfour \left( \adag_k \aphat - \ahat_k \apdag \right) \right]
\nonumber \\
&& - g'_- \left[ \sumkfour s_k \left( \adag_k \amhat - \ahat_k \amdag \right) \right], 
\label{eq:S}
\end{eqnarray}
where dimensionless coupling constants $g'_+$ and $g'_-$ are determined as
\begin{equation}
g'_+ 
= \frac{g_+}{\omega - \omega_+ - K}, 
\label{eq:g'+}
\end{equation}
\begin{equation}
g'_- 
= \frac{g_-}{\omega - \omega_- - K + \Km}.
\label{eq:g'-}
\end{equation}
The form of $\hat{S}$ has been determined to renormalize the coupling between the JPOs and the coupler into the JPOs as shown in Appendix~\ref{sec:effective_Hamiltonian}. 
To describe JPO $k$ in the frame rotating at half a pump frequency $\omegap{k}/2$
and the two modes of the coupler in the frame rotating at $\omega_+$ and $\omega_-$, respectively, 
we use 
\begin{equation}
\hat{U}_r
= e^{-i \sumkfour \Theta_k(t) \adag_k \ahat_k - i \omega_+ t \apdag \aphat - i \omega_- t \amdag \amhat},
\label{eq:U_r} 
\end{equation}
where $\Theta_k(t) = (\omegap{k} t + \theta_{p,k})/2$. The Hamiltonian then becomes $\Hhat' = \hat{U}_r^\dagger \hat{U}_g^\dagger \Hhat \hat{U}_g \hat{U}_r - i \hat{U}_r^\dagger \dot{\hat{U}}_r$. As shown in Appendix~\ref{sec:effective_Hamiltonian}, this transformation 
under the condition given by Eq.~(\ref{eq:pump_frequency}), the rotating-wave approximation, and truncation of higher order terms for $g'_\pm$ gives
\begin{widetext}
\begin{eqnarray}
\frac{\Hhat'}{\hbar}
=&& \sumkfour 
\left[
\Delta_k \adag_k \ahat_k 
- \frac{K'}{2} \adagtwo_k \ahattwo_k 
+ \frac{p}{2} \left( \adagtwo_k + \ahattwo_k \right)
\right]
\nonumber
\\
&&- \gfour \left( 
e^{i \sumkfour s_k \theta_{p,k} / 2} \adag_1 \adag_2 \ahat_3 \ahat_4 
+ e^{-i \sumkfour s_k \theta_{p,k} / 2} \ahat_1 \ahat_2 \adag_3 \adag_4 
\right)
- \sum_{k<l} 
\gtwo_{kl}
\adag_k \ahat_k \adag_l \ahat_l, 
\label{eq:Hhat'}
\end{eqnarray}
\end{widetext}
where
\begin{eqnarray}
\Delta_k 
=&& \omega \left(1 + g'^2_+ + g'^2_- - 4 g'^4_+ - 4 g'^4_- \right) 
\nonumber \\
&&+ K \left( 1 - 5 g'^2_+ -5 g'^2_- + 6 g'^4_+ + 6 g'^4_-  + g'^2_+ g'^2_- \right) 
\nonumber \\
&&- \omega_+ \left( g'^2_+ - 4 g'^4_+ \right)
- \omega_- \left( g'^2_- - 4 g'^4_- \right)
\nonumber \\
&&+ \Km \left( g'^2_- -2 g'^4_- \right) 
- \frac{\omegap{k}}{2}, 
\end{eqnarray}
\begin{eqnarray}
K' 
=&& \left( 1 - 2 g'^2_+ -2 g'^2_- + \frac{13}{6} g'^4_+ + \frac{13}{6} g'^4_- + 3 g'^2_+ g'^2_- \right) K
\nonumber \\
&&+ \frac{g'^4_-}{2} \Km, 
\end{eqnarray}
\begin{equation}
\gfour
= 2 g'^4_- \Km, 
\label{eq:gfour}
\end{equation}
and
\begin{equation}
\gtwo_{kl}
= \left( 2 g'^4_+ + 2 g'^4_- 
+ 4 g'^2_+ g'^2_- s_k s_l
 \right) K 
+ 2 g'^4_- \Km, 
\end{equation}
represented by the parameters in Eqs.~(\ref{eq:omega})--(\ref{eq:Kg-}), (\ref{eq:g'+}), and (\ref{eq:g'-}). 

Importantly, $\Hhat'$ describes the effective four-body coupling $\adag_1 \adag_2 \ahat_3 \ahat_4$ of JPOs with the coupling constant $\gfour$ and a phase factor $\exp(i\sumkfour s_k \theta_{p,k})$. 
The parameter dependence of $\gfour$ will be examined in the next section. 
The phase is composed of the constant phases $\theta_{p,k}$ of the pumps for each JPO
and costants $s_k$ given in Eq.~(\ref{eq:s}). 
We can control the magnitude of the four-body coupling and its sign by tuning $\theta_{p,k}$. 
Hence, the absolute value of $\gfour$ represents the maximum value of the four-body coupling constant. 
The condition for the pump frequencies [Eq.~(\ref{eq:pump_frequency})] makes the four-body coupling term stationary and prevents it from vanishing under the rotating-wave approximation as shown in Appendix~\ref{sec:effective_Hamiltonian}. The cross-Kerr term $\adag_k \ahat_k \adag_l \ahat_l$ survives as well. A previous study~\cite{S.Puri2017Jun} has shown that the cross-Kerr term only leads to a constant shift in energy if the amplitude of the coherent states is large and that it does not affect the result of quantum annealing. We leave the detailed analysis of this point as a future work. 

\section{Parameter dependence of the four-body coupling}
\label{sec:parameter-dependence}

\begin{table*}
\caption{\label{table:parameters}
Parameters discussed in Sec.~\ref{sec:parameter-dependence}.
The array branch and the smaller JJ branch are the two branches of the superconducting loop of the coupler [Fig.~\ref{fig:circuit}(a)].
}
\begin{ruledtabular}
\begin{tabular}{c l l}
Parameter & What the parameter means & Related figures and equations
\\
\hline
$C_J$ & Capacitance shunting the SQUID of a JPO & Fig.~\ref{fig:circuit}(b)
\\
$E_J$ & Josephson energy of each JJ of a JPO & Fig.~\ref{fig:circuit}(b)
\\
$\EJSigma$ & Static Josephson energy of a JPO & Eqs.~(\ref{eq:EJPhik}) and (\ref{eq:EJSigma})
\\
$I_c$ & Critical current of each JJ of a JPO &
\\
$\Phidc$ & Static offset of the external magnetic flux threading the SQUID of a JPO & Eq.~(\ref{eq:Phik})
\\
$\omega$ & Resonance frequency of a JPO & Eqs.~(\ref{eq:Hhat}) and (\ref{eq:omega})
\\
$C_g$ & Capacitance of the coupler & Fig.~\ref{fig:circuit}(a)
\\
$\ECgp$ & Charging energy of the coupler & Eq.~(\ref{eq:ECg'})
\\
$E_{J_g}$ & Josephson energy of each JJ in the array branch of the coupler & Fig.~\ref{fig:circuit}(a)
\\
$E_{J_g}^{(m)}$ & 
Coefficient of the $m$-th order term in couper's potential function & Eqs.~(\ref{eq:Ug}) and (\ref{eq:EJgm})
\\
$I_{cg}$ & Critical current of each JJ in the array branch of the coupler &
\\
$n$ & Number of JJs in series in the array branch of the coupler & Fig.~\ref{fig:circuit}(a)
\\
$\alpha$ & Area ratio of JJs between the array and smaller JJ branches of the coupler & Fig.~\ref{fig:circuit}(a)
\\
$\omega_-$ & Resonance frequency for coupler's mode $\amhat$ & Eqs.~(\ref{eq:Hhat}), (\ref{eq:omega-}), and (\ref{eq:omega-_circuit_param})
\\
$C$ & Coupling capacitance between a JPO and the coupler & Fig.~\ref{fig:circuit}(a)
\\
$\Omega$ & Effective detuning between a JPO and the coupler & Eq.~(\ref{eq:Omega}) 
\\
$g'_-$ & Dimensionless coupling constant of a JPO and the coupler & Eqs.~(\ref{eq:g'-}) and (\ref{eq:g'-_circuit_param})
\\
$\gfour$ & Four-body coupling constant of the JPOs & Eqs.~(\ref{eq:gfour}) and (\ref{eq:gfour_circuit_param})
\end{tabular}
\end{ruledtabular}
\end{table*}

We examine the parameter dependence of the four-body coupling constant $\gfour$ [Eq.(\ref{eq:gfour})] to find a way to increase its absolute value.
For convenience, 
the parameters discussed in this section are listed in Table~\ref{table:parameters}. 
We can rewrite the four-body coupling constant as
\begin{equation}
\gfour = \frac{2 g'^4_- \ECgp \EJgfour}{\hbar \EJgtwo},
\label{eq:gfour_rewritten}
\end{equation}
by substituting Eq.~(\ref{eq:Kg-}) into Eq.~(\ref{eq:gfour}). 
This expression shows that three factors are involved in $\gfour$. 
One is the dimensionless coupling constant $g'_-$ [Eq.~(\ref{eq:g'-})] between each JPO and the coupler. 
The second one is the charging energy $\ECgp$ of the coupler [Eq.~(\ref{eq:ECg'})]. 
The last one is the normalized amplitude of coupler's nonlinearity $\EJgfour/\EJgtwo$, 
where $\EJgtwo$ and $\EJgfour$ are the coefficients of the quadratic and quartic terms in coupler's potential function [Eq.~(\ref{eq:Ug})].
Note that $g'_-$ and $\EJgfour/\EJgtwo$ can be negative, 
but it does not matter, 
since the sign of $\gfour$ can be changed by tuning the constant phases $\theta_{p,k}$ of the pumps for each JPO. 
Although enhancing every factor increases $\gfour$, 
we cannot increase $g'_-$ much, 
since the coupling between JPOs and the coupler is assumed as a small perturbation as shown in, for example, Eq.~(\ref{eq:UdaggergakUg}). 
Therefore, we consider tuning $\ECgp$ and $|\EJgfour/\EJgtwo|$. 
The former, given in Eq.~(\ref{eq:ECg'}), can be increased by decreasing the capacitance $C_g$ of the coupler. 
The latter 
\begin{equation}
\left| \frac{\EJgfour}{\EJgtwo} \right| 
= \left| \frac{\left(1 - n^3 \alpha \right)}{n^2 \left( 1 - n\alpha \right)} \right|
\label{eq:EJg4_EJg2}
\end{equation}
can be increased by making $\alpha$ close to $1/n$, when $n \neq 1$, 
where $\alpha$ is the area ratio of JJs between the array and smaller JJ branches of the coupler, 
and $n$ is the number of JJs in series in the array branch. 
We do not set $\alpha$ to exactly $1/n$, 
because this setting leads to divergence of the zero-point fluctuations $\varphi_{g_{\pm}Z}$ in Eq.~(\ref{eq:varphik}) 
as derived from Eqs.~(\ref{eq:EJgm}), (\ref{eq:varphikZ}) and (\ref{eq:Zgpm}).
The divergence means our formulation is not valid for this setting.
When $n = 1$, $|\EJgfour/\EJgtwo|$ becomes unity for any $\alpha$, and this factor does not contribute to $\gfour$. Our generalized circuit with $n > 1$ enables $\gfour$ to be increased by tuning $\alpha$.

We have to consider that the above two factors can affect the first factor, 
i.e., the dimensionless coupling constant $g'_-$. 
To clarify this point, 
we represent the dimensionless coupling constant $g'_-$ and the four-body coupling constant $\gfour$ with circuit parameters. 
Using Eqs.~(\ref{eq:EJgm}), (\ref{eq:EC}), (\ref{eq:ECg'}), (\ref{eq:K}), (\ref{eq:omega-})--(\ref{eq:g-}), (\ref{eq:g'-}), and (\ref{eq:gfour}), 
we obtain 
\begin{equation}
g'_-
= \frac{\sqrt{\omega \omega_-} C}{4 \Omega \sqrt{C_J \left( C_g + C \right)}}, 
\label{eq:g'-_circuit_param}
\end{equation}
\begin{equation}
\gfour
= \frac{\left(\omega \omega_- \right)^2 C^4 e^2(1-n^3\alpha)}{\hbar \left(
4 \Omega
\right)^4 
C_J^2 \left( C_g + C \right)^3 n^2 (1-n\alpha)},
\label{eq:gfour_circuit_param}
\end{equation}
where we have introduced the effective detuning between a JPO and the coupler
\begin{equation}
\Omega
= \omega - \omega_-
- \frac{e^2}{2\hbar C_J} 
+ \frac{e^2\left(1 - n^3 \alpha \right)}{2\hbar(C_g + C)n^2 \left( 1 - n\alpha \right)}.
\label{eq:Omega}
\end{equation}
Note that 
a resonance frequency $\omega_-$ of the coupler is related to the two factors and depends on circuit parameters as
\begin{equation}
\omega_-
= \frac{1}{\hbar}\sqrt{\frac{4e^2}{C_g + C} \left( \frac{1}{n} - \alpha \right) \EJg}.
\label{eq:omega-_circuit_param}
\end{equation}
We can confirm a natural consequence that $\gfour$ becomes zero 
when we decrease the coupling capacitance $C$ between a JPO and the coupler to zero. 
Decreasing the capacitance $C_g$ shunting the coupler seems to increase $g'_-\propto (C_g + C)^{-1/2}$, 
but the increase is milder than $\gfour \propto (C_g + C)^{-3}$. 
This rough estimation assumes fixed $\omega_-$ and $\Omega$. 
However, they depend on $C_g$. 
The actual $C_g$ dependence of $g'_-$ and $\gfour$ is shown below in a specific setting. 
Hereafter, for simplicity, we fix $\Omega$ at an appropriate value by tuning $\omega_-$ with Josephson energy $\EJg$ of coupler's JJs. 
$\Omega$ can also be controlled by adjusting the resonance frequency $\omega$ of a JPO in situ via the external magnetic flux. 
Note that 
$\omega$ [Eq.~(\ref{eq:omega})] depends on the static Josephson energy $\EJSigma$ of the JPO, 
which is controlled by the external magnetic flux [Eq.~(\ref{eq:EJSigma})].
As discussed below, 
$\omega$ works as the second control knob in setting $\Omega$, 
compensating for unavoided errors in $\omega_-$ after the device fabrication. 

\begin{figure}
\includegraphics[width=\columnwidth]{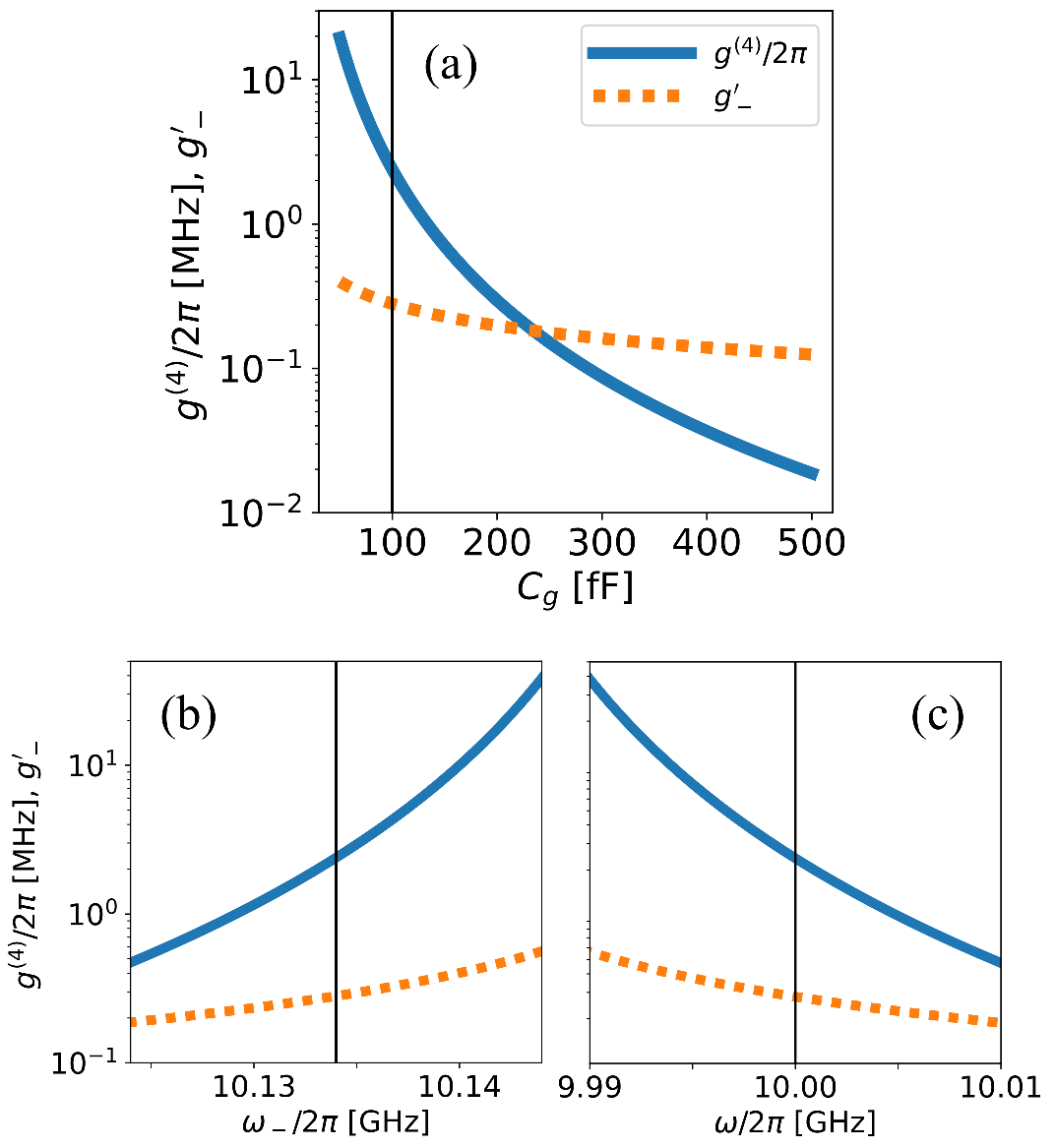}%
\caption{\label{fig:g4_g'-_Cg} (a) Four-body coupling constant $\gfour$ [Eq.~(\ref{eq:gfour_circuit_param})] and effective coupling constant $g'_-$ [Eq.~(\ref{eq:g'-_circuit_param})] between each JPO and the coupler as a function of capacitance $C_g$ of the coupler [Fig.~\ref{fig:circuit}]. 
The other parameters are set as follows: $n = 1$, $\alpha = 0$, $\omega = 2\pi \times 10$~GHz, $C_J =$ 500~fF, $C =$ 0.5~fF, and $\Omega = 2\pi \times 20$~MHz [Eq.~(\ref{eq:Omega})]. 
See Table~\ref{table:parameters} for what each parameter means.
$\omega_-$ is set to satisfy Eq.~(\ref{eq:Omega}) for this setup as a function of $C_g$. 
The vertical line indicates $C_g = 100$~fF, which is used in (b), (c), and Fig.~\ref{fig:g4_alpha_n}. 
(b) $\gfour$ and $g'_-$ as a function of frequency $\omega_-$ of the coupler. 
$n$, $\alpha$, $\omega$, $C_J$, and $C$ are the same as (a), but $\Omega$, defined by Eq.~(\ref{eq:Omega}), varies in accordance with $\omega_-$. $C_g$ is set to 100~fF. 
The vertical line indicates the value of $\omega_-/2\pi$ used in (a) for $C_g = 100$~fF. (c) $\gfour$ and $g'_-$ as a function of frequency $\omega$ of a JPO instead of $\omega_-$ in (b). 
$\omega_-$ is the same as (a) for $C_g = 100$~fF. The vertical line indicates the value of $\omega/2\pi$ used in (a). (b) and (c) show how changes in $\omega_-$ and $\omega$ from the values in (a) affect $\gfour$ and $g'_-$.}
\end{figure}

As mentioned above, we only consider the second and third factors. 
We first examine the impact of the second factor, i.e., coupler's charging energy $\ECgp$ 
that increases as the capacitance $C_g$ of the coupler decreases. 
Figure \ref{fig:g4_g'-_Cg}(a) shows the four-body coupling constant $\gfour$ and the dimensionless coupling constant $g'_-$ 
as a function of $C_g$ for the following parameter setting. 
See Table~\ref{table:parameters} for what each parameter means.
To focus on the $C_g$ dependence, 
we set $n$ and $\alpha$ to 1 and 0, respectively. 
Other parameters are $\omega = 2\pi \times 10$~GHz, $C_J = $500~fF, and $C = 0.5$~fF. 
For simplicity, $\omega_-$ is assumed to be finely tuned with $\EJg$
so that $\Omega$ [Eq.~(\ref{eq:Omega})] is $2\pi \times 20$~MHz, 
where $\EJg$ needs to be increased roughly proportional to $C_g$. 
This figure demonstrates that we can increase $\gfour$ by decreasing $C_g$ under fixed $\Omega$, while $g'_-$ does not become so large. In other words, we can enhance the second factor $\ECgp$ while keeping the first factor $g'_-$ small as a perturbative parameter. 
For instance, we have $\gfour \simeq 2\pi \times$2.4~MHz and $g'_- \simeq 0.28$ for $C_g =$100~fF. 
The obtained value of four-body coupling constant is based only on the leading term for $g'_-$, i.e., Eq.~(\ref{eq:gfour}). 
The main contribution of the next higher order terms is $-8K_- g'^6_-/3$. 
Its ratio to $\gfour$ [Eq.~(\ref{eq:gfour})] is -0.21 for the above setting. 
Hence the contribution of higher order terms is smaller than that of the leading term, but the above ratio might not satisfy the required accuracy in some cases. We can obtain a more accurate four-body coupling constant by taking into account the higher order terms, but we do not pursue it here. Quantitatively precise computation is outside the scope of this paper.

We have assumed that 
a resonance frequency $\omega_-$ of the coupler is finely tuned, 
but it would have some fabrication errors in reality. 
Figure~\ref{fig:g4_g'-_Cg}(b) shows how $\gfour$ and $g'_-$ are affected by errors in $\omega_-$. $n$, $\alpha$, $\omega$, $C_J$, and $C$, are the same as those in Fig.~\ref{fig:g4_g'-_Cg}(a), but $C_g$ is set to 100 fF. The vertical line in Fig.~\ref{fig:g4_g'-_Cg}(b) indicates $\omega_-$ used in Fig.~\ref{fig:g4_g'-_Cg}(a). Errors in $\omega_-$ can cause large changes in $\gfour$, while $g'_-$ shows weaker dependence. The decrease in $\gfour$ induced by errors in $\omega_-$ is avoidable by adjusting $\omega$, 
because $\omega$ similarly affects $\gfour$ as shown in Fig.~\ref{fig:g4_g'-_Cg}(c). 
Note that $\omega$ is independent of $C_g$ 
and can easily be tuned in situ 
by the magnetic flux threading the SQUID of a JPO 
as shown in Eqs.~(\ref{eq:EJSigma}) and (\ref{eq:omega}). 
Thus, we can compensate for the errors in $\omega_-$ by tuning $\omega$. 

\begin{figure}
\includegraphics[width=\columnwidth]{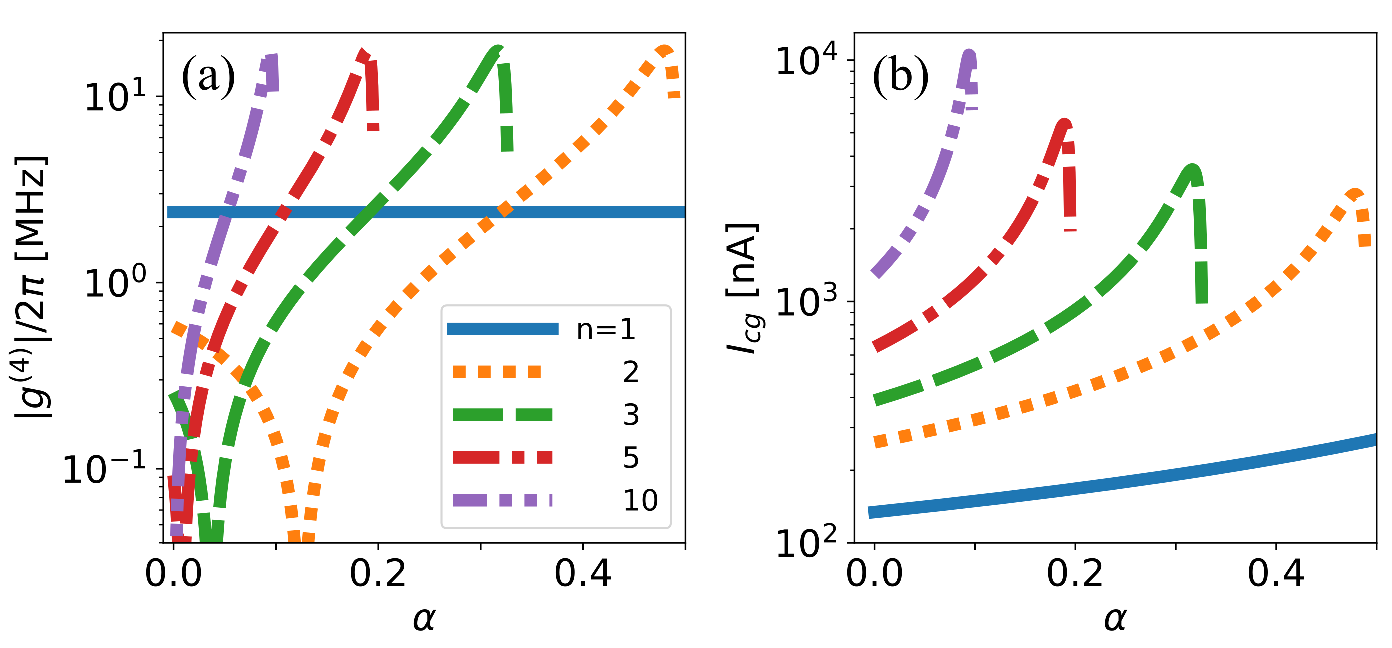}
\caption{\label{fig:g4_alpha_n} (a) Four-body coupling constant $\gfour$ [Eq.~(\ref{eq:gfour_circuit_param})] and (b) the critical current $I_{cg}$ of each JJ in a branch of the coupler required to set $\Omega = 2\pi \times 20$~MHz [Eq.~(\ref{eq:Omega})] as a function of $\alpha$ for $n = 1, 2, 3, 5, 10$, 
where $\alpha$ is the area ratio of JJs in the two branches of coupler's superconducting loop, 
and $n$ is the number of JJs in series in a branch [Fig.~\ref{fig:circuit}(a)].
The other parameters are the same as those for Fig.~\ref{fig:g4_g'-_Cg}~(a), except for $C_g$ set to 100 fF.}
\end{figure}

We move to the third factor, i.e., normalized amplitude of the coupler's nonlinearity $|\EJgfour/\EJgtwo|$ 
that depends on $n$ and $\alpha$ [Eq.~(\ref{eq:EJg4_EJg2})], 
where $n$ is the number of JJs in series in the array branch of the coupler, 
and $\alpha$ is the area ratio of JJs between the array and smaller JJs branches [Fig.~\ref{fig:circuit}(a)]. 
The absolute value $|\gfour|$ of the four-body coupling constant 
as a function of $\alpha$ for $n = 1, 2, 3, 5$, and 10 is shown in Fig.~\ref{fig:g4_alpha_n}(a). 
The other parameters are the same as those in Fig.~\ref{fig:g4_g'-_Cg}(a), except for the capacitance $C_g$ of the coupler set to 100~fF. 
$|\EJgfour/\EJgtwo|$ for $n = 1$ does not depend on $\alpha$, leading to the constant $|\gfour|$ drawn as a straight line in the figure. 
$|\gfour|$ for $n > 1$ varies with $\alpha$ due to the $\alpha$ dependence of $|\EJgfour/\EJgtwo|$. 
Starting from $\alpha = 0$, $\gfour$ decreases with increasing $\alpha$ 
and becomes negative at $\alpha =n^{-3}$. 
Then $|\gfour|$ increases rapidly with $\alpha$ 
and exceeds that for $n = 1$. 
Larger $n$ makes $|\gfour|$ increase more steeply. 
$|\gfour|$ decreases again only in a small window of $\alpha$ close to $1/n$. 
This is because a resonance frequency $\omega_-$ of the coupler in the numerator of $\gfour$ [Eq.~(\ref{eq:gfour_circuit_param})] for fixed $\Omega$ [Eq.~(\ref{eq:Omega})] becomes smaller as $\alpha$ increases. 
Note that increasing $\alpha$ close to $1/n$ largely decreases the fourth term in Eq.~(\ref{eq:Omega}), 
and thus the second term $\omega_-$ is set to a smaller value to fix $\Omega$. 
Although $|\gfour|$ decreases in this range, it is still larger than that for $n=1$. 
This result means that 
our coupler can induce a stronger four-body coupling than that with a single JJ, i.e., $n = 1$ and $\alpha = 0$, by tuning $\alpha$ for $n > 1$.

In Fig.~\ref{fig:g4_alpha_n}(a), 
we have assumed $\Omega$ [Eq.~(\ref{eq:Omega})] to be a constant, 
although it depends on $\alpha$. 
$\Omega$ can be fixed by adjusting Josephson energy $\EJg$ of coupler's JJs 
or their critical currents $I_{cg} = (2e/\hbar) \EJg$ 
to tune $\omega_-$ 
and conserve the sum of the second and fourth terms in Eq.~(\ref{eq:Omega}). 
Figure~\ref{fig:g4_alpha_n}(b) shows required $I_{cg}$ to set $\Omega = 2\pi \times 20$~MHz as a function of $\alpha$ for given $n$. 
The other parameters are the same as those in Fig.~\ref{fig:g4_alpha_n}(a). 
We find that required $I_{cg}$ rapidly increases with $\alpha$, 
although it starts to decline at $\alpha$ close to $1/n$ as $|\gfour|$. 
This fact suggests that 
errors in fabrication of $\alpha$ cause errors in estimation of the required $I_{cg}$. 
The caused errors tend to increase with $n$,
because larger $n$ makes the required $I_{cg}$ increase more steeply.
In other words, smaller $n$ is more tolerant of $\alpha$.
These errors result in $\Omega$ largely deviating from the expected one.
Although such a deviation of $\Omega$ may result in a decrease of $\gfour$, 
it can be mitigated by adjusting $\omega$. 

\begin{figure}
\includegraphics[width=0.7\columnwidth]{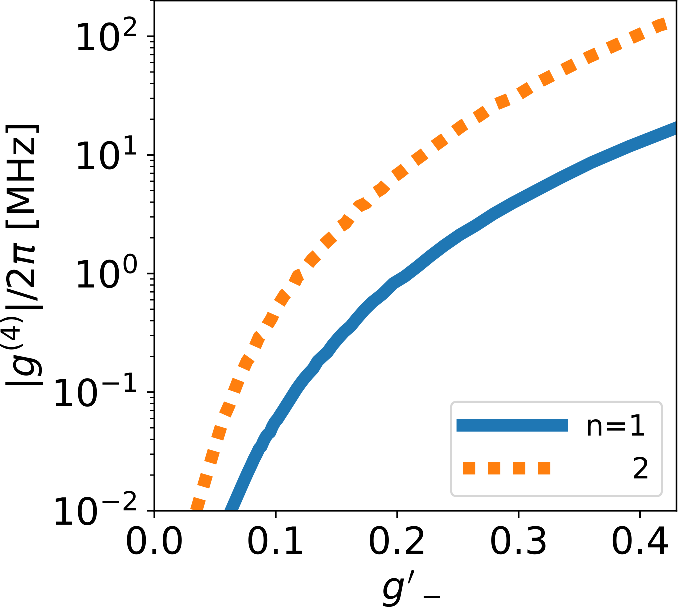}
\caption{\label{fig:g4_gmp} 
			Maximum absolute value $|\gfour|$ of the four-body coupling constant [Eq.~(\ref{eq:gfour_circuit_param})] 
			as a function of the effective coupling constant $g'_-$ [Eq.~(\ref{eq:g'-_circuit_param})] for $n = 1, 2$, 
			where $n$ is the number of JJs in series in a branch of the coupler
			The parameters $C_J$, $C_g$, $C$, $\omega_-$, and $\alpha$ are optimized 
			to maximize $|\gfour|$ under the constraints described in main text.
			See Table~\ref{table:parameters} for what each parameter means.
			The range of $g'_-$ is determined 
			so that the absolute value of the main contribution $-8K_- g'^6_-/3$ of the next higher order terms 
			is smaller than half of $|\gfour|$, 
			which is based only on the leading term for $g'_-$. 
			Unlike Fig.~\ref{fig:g4_alpha_n}, 
			the corresponding curves for $n = 3, 5, 10$ are not shown, 
			because they almost overlap with the curve for $n=2$ 
			and are hard to distinguish.}
\end{figure}

It is worth calculating the maximum absolute value $|\gfour|$ of the four-body coupling constant 
as a function of the dimensionless coupling constant $g'_-$ for a given $n$ 
to evaluate the total enhancement of the four-body coupling with our coupler.
We optimize $C_J$, $C_g$, $C$, $\omega_-$, and $\alpha$ 
with fixed $\omega$ at $2\pi \times 10$~GHz 
to maximize $|\gfour|$ in the following parameter ranges: 
50~fF~$\le C_J \le$~500~fF, 
50~fF~$\le C_g \le$~250~fF, 
0.1~fF~$\le C \le 0.01 C_J \le$~5~fF, 
8~GHz~$\le \omega_- \le$~12~GHz, 
$0 \le \alpha \le 0.99/n$.
See Table~\ref{table:parameters} for what each parameter means.
$\EJSigma$ and $E_{J_g}$ are determined to be consistent with these parameters 
and Eqs.~(\ref{eq:EC}), (\ref{eq:omega}), and (\ref{eq:omega-_circuit_param}).
We set $\Phidc$ = $\Phi_0 /4$ in Eq.~(\ref{eq:EJSigma}) 
and obtain the Josephson energy $E_J$ of each JJ in the JPOs. 
We impose constraints on the critical currents $I_c = (2e/\hbar) E_J$ and $I_{cg} = (2e/\hbar) E_{J_g}$ as 
100~nA $\le I_c$, $I_{cg} \le$ 10000~nA.
We use only the parameter settings that satisfy these constraints.
Figure \ref{fig:g4_gmp} shows the obtained maximum $|\gfour|$ as a function of $g'_-$ for $n = 1, 2$.
$|\gfour|$ is valid for a small $g'_-$, 
since it is based only on the leading term for $g'_-$.
The range of $g'_-$ in the figure is determined so that 
the absolute value of the main contribution $-8K_- g'^6_-/3$ 
of the next higher order terms is smaller than half of $|\gfour|$.
The maximum $|\gfour|$ for $n = 2$ is always approximately ten times larger 
than that for $n = 1$ in the range.
This result shows that 
our coupler can enhance the four-body coupling of JPOs 
better than the single JJ coupler
for given effective coupling constants between the JPOs and the coupler. 
The corresponding curves for $n = 3, 5, 10$ are not shown, 
because they almost overlap with the curve for $n=2$.
While the cause of this similarity has not been identified,  
the similarity suggests that
two JJs in series is enough to maximize the effectiveness of our coupler.
Furthermore,  
$n = 2$ is better than larger $n$ in the sense that the tolerance of $\alpha$ is larger 
as mentioned above.

\section{Conclusion}
\label{sec:conclusion}

We have proposed a superconducting circuit [Fig.~\ref{fig:circuit}] for the four-body interaction of Josephson parametric oscillators (JPOs). 
The coupler has a superconducting loop consisting of two parallel branches and has additional control knobs $n$ and $\alpha$, 
where $n$ is the number of Josephson junctions (JJs) in series in one of the branches, 
and $\alpha$ is the area ratio of JJs in the two branches. 
The four-body coupling has been derived under some assumptions about the circuit parameters and the rotating-wave approximation. 
The obtained four-body coupling constant is composed of three factors: 
the coupling constant of each JPO and the coupler, coupler's charging energy, and the normalized amplitude of coupler's nonlinearity. 
The four-body coupling constant can be increased by enhancing the latter two factors, i.e., decreasing capacitance of the coupler and setting $\alpha$ close to $1/n$ with $n > 1$. 
We also need to carefully set the frequency of the coupler to prevent the first factor from increasing too much. 
Unavoided errors in the frequency of the coupler can also cause drastic changes in the four-body coupling constant, 
but they can be compensated for by adjusting the frequency of JPOs, 
which is easily tuned with the external magnetic flux. 
Our result predicts that 
tuning parameters in accordance with the above guideline leads to stronger four-body couplings than that with the coupler composed of a single JJ where $n = 1$ and $\alpha = 0$.

The obtained analytical expression of coupling constant clarifies roles of circuit parameters in the four-body coupling of JPOs. 
In particular, the expression shows the above three factors and facilitates the advancement of insight into the four-body coupling. 
The expression also gives guidelines for parameter setting. 
Although the investigated circuit is simple and would have some differences from actual experimental settings, 
these results should be useful to implement the four-body interaction as a starting point. 
Detailed quantitative investigations for such actual circuits will become an important task when the four-body interaction based on our theoretical proposal is physically realized. 

\begin{acknowledgments}
The authors thank Y. Ye, K. P. O'Brien, Y. Kawakami, A. Yamaguchi, and T. Yamaji for valuable discussions. 
This paper is based on results obtained from Project No. JPNP16007 commissioned by the New Energy and Industrial Technology Development Organization (NEDO), Japan. 
\end{acknowledgments}

\appendix

\begin{widetext}

\section{Effective Josephson energy}
\label{sec:effective_Josephson_energy}

The potential function $V_k(\phi_k)$ of JPO $k$ shown in Fig.~\ref{fig:circuit}(b) is represented as 
the sum of the potential functions for the two Josephson junctions 
in the SQUID of JPO $k$, 
\begin{eqnarray}
V_k(\phi_k)
=&& - E_J \cos \left( \phi_k + \pi \frac{\Phi_k}{\Phi_0} \right)
- E_J \cos \left( -\phi_k + \pi \frac{\Phi_k}{\Phi_0} \right)
\nonumber \\
=&& - 2E_J \cos \left( \pi \frac{\Phi_k}{\Phi_0} \right) 
\cos \phi_k,  
\end{eqnarray}
where $\phi_k$ is a node flux in Fig.~\ref{fig:circuit}(a), 
and $\Phi_0$ is the flux quantum.
We have assumed that the two Josephson junctions have the same Josephson energy $E_J$.
In the first line 
the phase $2\pi \Phi_k/\Phi_0$ associated with the magnetic flux $\Phi_k$ has been equally divided into the two phases
to appropriately incorporate the time-dependent flux~\cite{X.You2019}. 
The magnetic flux is modulated as Eq.~(\ref{eq:Phik}).
This gives
\begin{eqnarray}
\cos \left( \pi \frac{\Phi_k}{\Phi_0} \right)
=&& \cos \left( \pi \frac{\Phidc}{\Phi_0} \right) 
\cos \left[ \pi \frac{\delta\Phi}{\Phi_0} \cos \left( \omega_{p,k} t + \theta_{p,k} \right) \right]
- \sin \left( \pi \frac{\Phidc}{\Phi_0} \right) 
\sin \left[ \pi \frac{\delta\Phi}{\Phi_0} \cos \left( \omega_{p,k} t + \theta_{p,k} \right) \right]
\nonumber \\
=&& \cos \left( \pi \frac{\Phidc}{\Phi_0} \right) 
- \sin \left( \pi \frac{\Phidc}{\Phi_0} \right) 
\pi \frac{\delta\Phi}{\Phi_0} \cos \left( \omega_{p,k} t + \theta_{p,k} \right)
+ (\text{rapidly oscillating terms}), 
\end{eqnarray}
where the rapidly oscillating terms include terms for higher frequencies, 
e.g., $2\omega_{p, k}$.
Thus the above potential function can be represented as 
$V_k(\phi_k) = - E_J(\Phi_k) \cos \phi_k$ 
with the effective Josephson energy 
\begin{equation}
E_J(\Phi_k)
= -2E_J \cos \left( \pi \frac{\Phidc}{\Phi_0} \right)  
+ 2E_J \sin \left( \pi \frac{\Phidc}{\Phi_0} \right) 
\pi \frac{\delta\Phi}{\Phi_0} \cos \left( \omega_{p,k} t + \theta_{p,k} \right), 
\label{eq:EJPhik_2}
\end{equation}
where the rapidly oscillating terms have been ignored.
This expression gives the parameters in Eq.~(\ref{eq:EJPhik}) as Eqs.~(\ref{eq:EJSigma}) and (\ref{eq:deltaEJ}).

\section{Classical Hamiltonian}
\label{sec:classical_Hamiltonian}

The capacitance matrix in the Lagrangian [Eq.~(\ref{eq:L})] is
\begin{equation}
\Cmat
=
\begin{pmatrix} 
C_J+C & 0 & 0 & 0 & -C & 0 \\
0 & C_J+C & 0 & 0 & -C & 0 \\
0 & 0 & C_J+C & 0 & 0 & -C \\
0 & 0 & 0 & C_J+C & 0 & -C \\
-C & -C & 0 & 0 & C_g + 2C & -C_g \\
0 & 0 & -C & -C & -C_g & C_g+2C
\end{pmatrix}.
\end{equation}
The inverse matrix consists of the following elements:
\begin{align}
\Cmatinv{11}
= \Cmatinv{22}
= \Cmatinv{33}
= \Cmatinv{44}
=& \frac{1}{C_J + C} 
\left[
1 + \frac{C}{4C_J} + \frac{C^2}{4(C_J C + C C_g + C_g C_J)} 
\right], 
\\
\left( \Cmat^{-1} \right)^{-1}_{12} 
= \left( \Cmat^{-1} \right)^{-1}_{34} 
=& \frac{C}{C_J + C} 
\left[
\frac{1}{4C_J} + \frac{C}{4(C_J C + C C_g + C_g C_J)} 
\right], 
\\
\Cmatinv{13} 
= \Cmatinv{14} 
= \Cmatinv{23} 
= \Cmatinv{24} 
=& \frac{C C_g}{4C_J (C_J C + C C_g + C_g C_J)}, 
\\
\Cmatinv{15} 
= \Cmatinv{25} 
= \Cmatinv{36} 
= \Cmatinv{46} 
=&\frac{1}{4C_J} + \frac{C}{4(C_J C + C C_g + C_g C_J)}, 
\\
\Cmatinv{16} 
= \Cmatinv{26} 
= \Cmatinv{35} 
= \Cmatinv{45} 
=& \frac{1}{4C_J} - \frac{C}{4(C_J C + C C_g + C_g C_J)} 
\\
\Cmatinv{55} 
= \Cmatinv{66} 
=&\frac{1}{4C} + \frac{1}{4C_J} + \frac{C_J + C}{4(C_J C + C C_g + C_g C_J)}, 
\\
\Cmatinv{56} 
=&\frac{1}{4C} + \frac{1}{4C_J} - \frac{C_J + C}{4(C_J C + C C_g + C_g C_J)}.
\end{align}
The other elements are obtained by using $(\Cmat^{-1})_{ij} = (\Cmat^{-1})_{ji}$.
The Hamiltonian in Eq.~(\ref{eq:H_formal}) then reads
\begin{eqnarray}
\mathcal{H}
=&& \frac{1}{2} \Cmatinv{11} \sumkfour q_k^2
+ \Cmatinv{12} \left(q_1 q_2 + q_3 q_4 \right) 
+ \Cmatinv{13} \left( q_1 + q_2 \right) \left(q_3 + q_4 \right) 
\nonumber \\
&& + \Cmatinv{15} \left( q_1 q_{g_1} + q_2 q_{g_1} + q_3 q_{g_2} + q_4 q_{g_2} \right) 
+ \Cmatinv{16} \left( q_1 q_{g_2} + q_2 q_{g_2} + q_3 q_{g_1} + q_4 q_{g_1} \right)
\nonumber \\
&& + \frac{1}{2} \Cmatinv{55} \left( q_{g_1}^2 + q_{g_2}^2 \right) 
+ \Cmatinv{56} q_{g_1} q_{g_2} 
+V\left(\bm{\phi}\right) 
\nonumber \\
=&& \frac{1}{2} \Cmatinv{11} \sumkfour q_k^2
+ \Cmatinv{12} \left(q_1 q_2 + q_3 q_4 \right) 
+ \Cmatinv{13} \left( q_1 + q_2 \right) \left(q_3 + q_4 \right) 
\nonumber \\
&& + \left[ \Cmatinv{15} + \Cmatinv{16} \right] \sumkfour  q_k q_{g_+} 
+ \left[ \Cmatinv{15} - \Cmatinv{16} \right] \sumkfour  s_k q_k q_{g_-} 
\nonumber \\
&& + \left[ \Cmatinv{55} + \Cmatinv{56} \right] q_{g_+}^2  
+ \left[ \Cmatinv{55} - \Cmatinv{56} \right] q_{g_-}^2  
+V\left(\bm{\phi}\right) 
\nonumber \\
=&& \frac{1}{2(C_J + C)} 
\left[
1 + \frac{C}{4C_J} + \frac{C^2}{4(C_J C + C C_g + C_g C_J)} 
\right] 
\sumkfour q_k^2 
\nonumber \\
&& + \frac{C}{C_J + C} 
\left[
\frac{1}{4C_J} + \frac{C}{4(C_J C + C C_g + C_g C_J)} 
\right] 
(q_1 q_2 + q_3 q_4) 
+ \frac{C C_g}{4C_J (C_J C + C C_g + C_g C_J)} (q_1 + q_2) (q_3 + q_4) 
\nonumber \\
&& + \frac{1}{2C_J} \sumkfour  q_k q_{g_+} 
+ \frac{C}{2(C_J C + C C_g + C_g C_J)} \sumkfour  s_k q_k q_{g_-} 
\nonumber \\
&& + \left[ \frac{1}{2C} + \frac{1}{2C_J} \right] q_{g_+}^2  
+ \frac{C_J + C}{2(C_J C + C C_g + C_g C_J)} q_{g_-}^2  
+V(\bm{\phi}), 
\end{eqnarray}
where $q_{g\pm} = (q_{g_1} \pm q_{g_2})/2$, 
$s_1 = s_2 = 1$, and $s_3 = s_4 = -1$.
The charging-energy part normalized 
by an energy scale $e^2/2C_J$, 
where $e$ denotes the elementary charge, 
is written as
\begin{eqnarray}
&&\frac{2C_J}{e^2} \left[ \mathcal{H} - V(\bm{\phi}) \right]
\nonumber \\
&&= \sumkfour \frac{q_k^2}{e^2} + \sumkfour  \frac{q_k q_{g_+}}{e^2} 
+ \frac{C}{C_g + C} \sumkfour  s_k \frac{q_k q_{g_-}}{e^2}
+ \left( \frac{C_J}{C} + 1 \right) \frac{q_{g_+}^2}{e^2}
+ \left[ \frac{C_J}{C_g + C} + \left( \frac{C}{C_g + C} \right)^2 \right] \frac{q_{g_-}^2}{e^2}
+ O\left( \frac{C}{C_J} \right).
\end{eqnarray}
Interactions between JPOs such as $q_1 q_2$ or $q_1 q_3$ are included 
in the last term $O(C/C_J)$.
Hence, those interactions are suppressed when $C \ll C_J$.
Ignoring the terms of $O(C/C_J)$ leads to Eq.~(\ref{eq:H}).

\section{Effective Hamiltonian}
\label{sec:effective_Hamiltonian}

We calculate $\hat{U}_r^\dagger \hat{U}_g^\dagger \Hhat \hat{U}_g \hat{U}_r - i \hat{U}_r^\dagger \dot{\hat{U}}_r$ to obtain the effective Hamiltonian $\Hhat'$ [Eq.~(\ref{eq:Hhat'})] derived from $\Hhat$ [Eq.~(\ref{eq:Hhat})], where $\hat{U}_g$ and $\hat{U}_r$ are given by Eqs.~(\ref{eq:U_g})--(\ref{eq:U_r}). Using these unitary operators, we transform the couplings between JPOs and the coupler into direct couplings between JPOs. 
The transformation is based on the Schrieffer-Wolf (SW) transformation~\cite{S.Bravyi2011}. 
To obtain the effective four-body coupling with the SW transformation, 
we need to find a unitary operator that diagonalizes the system up to the third order of dimensionless coupling constants $g'_{\pm}$ and derives the four-body coupling as the lowest order of non-diagonal terms. However, such a unitary operator is difficult to find for our system because of its nonlinearity and time dependency. Thus, we do not exactly perform the SW transformation but partially execute it with a rather simple unitary operator $\hat{U}_g$ and the rotating-wave approximation in the rotating frame mapped by $\hat{U}_r$. We only tune $g'_\pm$ in $\hat{S}$ [Eq.~(\ref{eq:S})] so that the coupling terms between JPOs and the coupler in $\Hhat$, which are represented by
\begin{equation}
\frac{\Hhat_g}{\hbar}
= \sumkfour \left[
g_+ \left( \adag_k \aphat + \ahat_k \apdag \right) 
+ g_- s_k \left( \adag_k \amhat + \ahat_k \amdag \right)
\right],
\label{eq:Hhatg}
\end{equation} 
are cancelled out by the transformation. 
$\Hhat$ also has other coupling terms, e.g., $g_+ \adag_1 \apdag$. 
They would not be cancelled out, 
but they will have rapidly oscillating phases after the transformation and can be ignored under the rotating-wave approximation. 
The coupling terms in $\Hhat_g$ also oscillate, 
but we do not ignore them, 
because it has a lower frequency.
This transformation is expected to incorporate the most dominant effects of the coupler into JPOs. The transformation derives the couplings between JPOs including the four-body one but also the nonlinear one between JPOs and the coupler. 
Most of the derived coupling terms have rapidly oscillating phases and can be ignored, while the four-body coupling term of JPOs has no oscillating phase under the condition of pump frequencies set by Eq.~(\ref{eq:pump_frequency}). The resulting $g'_\pm$ is given by Eqs.~(\ref{eq:g'+}) and (\ref{eq:g'-}). We describe how $g'_\pm$ is determined below.

First, we calculate 
\begin{equation}
\hat{U}_g^\dagger \ahat_k \hat{U}_g
= \ahat_k + [\ahat_k, \hat{S}] + \frac{1}{2} [ [\ahat_k, \hat{S}], \hat{S}] 
+ \frac{1}{6} [ [ [\ahat_k, \hat{S}], \hat{S}], \hat{S}] + \dots
\end{equation} 
for $k = 1, 2, 3, 4, g_+$, and $g_-$. The commutation relations
\begin{equation}
[\ahat_k, \hat{S}] 
= - g'_+ \aphat - s_k g'_- \amhat
\ \ \ 
\text{for } 
k = 1, 2, 3, 4, 
\end{equation}
\begin{equation}
[\aphat, \hat{S}]
= g'_+ \sumkfour \ahat_k, 
\ \ \ 
\text{and}
\ \ \ 
[\amhat, \hat{S}]
= g'_- \sumkfour s_k \ahat_k,  
\end{equation}
where $s_1 = s_2 = 1$ and $s_3 = s_4 = -1$ [Eq.~(\ref{eq:s})], give
\begin{eqnarray}
&&\hat{U}_g^\dagger \ahat_k \hat{U}_g
\nonumber \\
=&& \ahat_k 
- \left( g'_+ - \frac{2}{3} g'^3_+ \right) \aphat 
- s_k \left( g'_- - \frac{2}{3} g'^3_- \right) \amhat 
- \left( \frac{g'^2_+}{2} - \frac{g'^4_+}{6} \right) \sumlfour \ahat_l 
- s_k \left( \frac{g'^2_-}{2} - \frac{g'^4_-}{6} \right) \sumlfour s_l \ahat_l 
+ O (g'^5_\pm), 
\label{eq:UdaggergakUg}
\end{eqnarray}
\begin{eqnarray}
\hat{U}_g^\dagger \aphat \hat{U}_g
=&& \left( 1 - 2g'^2_+ + \frac{2}{3} g'^4_+ \right) \aphat
+ \left( g'_+ - \frac{2}{3} g'^3_+ \right) \sumkfour \ahat_k 
+ O(g'^5_\pm), 
\end{eqnarray}
and
\begin{eqnarray}
\hat{U}_g^\dagger \amhat \hat{U}_g
=&& \left( 1 - 2g'^2_- + \frac{2}{3} g'^4_- \right) \amhat
+ \left( g'_- - \frac{2}{3} g'^3_- \right) \sumkfour s_k \ahat_k 
+ O(g'^5_\pm), 
\end{eqnarray}
where $O(g'^5_\pm)$ represents terms of products including five or more $g'_+$s or $g'_-$s. We also calculate the unitary transformation with $\hat{U}_r$ using
\begin{equation}
\hat{U}_r^\dagger \ahat_k \hat{U}_r
= e^{-i\Theta_k(t)} \ahat_k, 
\ \ \ 
\Theta_k(t) 
= \frac{\omegap{k} t + \theta_{p,k}}{2}, 
\ \ \
\text{for } k = 1,2,3, 4, 
\label{eq:UrdaggerakUr}
\end{equation}
\begin{equation}
\hat{U}_r^\dagger \aphat \hat{U}_r
= e^{-i\omega_+ t} \aphat, 
\ \ \ 
\text{and}
\ \ \ 
\hat{U}_r^\dagger \amhat \hat{U}_r
= e^{-i\omega_- t} \amhat.
\end{equation}
$\Hhat'$ also has 
\begin{eqnarray}
- i \hat{U}_r^\dagger \dot{\hat{U}}_r
= - \sumkfour \frac{\omega_{p,k}}{2} \adag_k \ahat_k - \omega_+ \apdag \aphat - \omega_- \amdag \amhat.
\label{eq:UrdaggerUrdot}
\end{eqnarray}

Next, we expand the unitary transformation of the terms in $\Hhat$ [Eq.~(\ref{eq:Hhat})] using the above relations. We consider up to the first order of $g'_\pm$ to find appropriate $g'_\pm$ for cancelling out the coupling terms in $\Hhat_g$ [Eq.~(\ref{eq:Hhatg})]. For example, $\adag_k \ahat_k$ is expanded as
\begin{eqnarray}
\Utrans{\adag_k \ahat_k}
=&& \adag_k \ahat_k 
- g'_+ \left( e^{i\Thetakpt}\adag_k \aphat + \hc \right)
- s_k g'_- \left( e^{i\Thetakmt} \adag_k \amhat + \hc \right) 
+ O(g'^2_\pm), 
\end{eqnarray}
\begin{equation}
\Thetakpmt = \Theta_k(t) - \omega_\pm t, 
\end{equation}
where $\hc$ denotes the Hermitian conjugate of its preceding term. $\Thetakpmt$ is the oscillating phase that will be attained by the coupling terms between JPOs and the coupler in $\Hhat_g$ [see Eqs.~(\ref{eq:unitary_trans_adagkaphat}) and (\ref{eq:unitary_trans_adagkamhat})]. Hereafter, we focus on the terms with $e^{i \Thetakpmt}$. The same treatment of $( \ahat_k + \adag_k )^4$ leads to
\begin{eqnarray}
\Utrans{\left( \ahat_k + \adag_k \right)^4} 
=&& 6 \adagtwo_k \ahat_k^2 - 12\adag_k \ahat_k + 3
\nonumber \\
&&-12 g'_+ \left[ 
e^{i \Thetakpt} \adag_k \left( \adag_k \ahat_k + 1 \right) \aphat + \hc
\right] 
-12 s_k g'_- \left[ 
e^{i\Thetakmt} \adag_k \left( \adag_k \ahat_k + 1 \right) \amhat + \hc 
\right] 
\nonumber \\
&&+ (\text{other oscillating terms}) + O(g'^2_\pm), 
\end{eqnarray}
where other oscillating terms contain terms oscillating with other frequencies than $\Thetakpmt$, e.g., $e^{i[3\Theta_k(t)- \omega_+ t]} \hat{a}_k^{\dagger 3} \aphat$. The other terms in $\Hhat$ are transformed as follows:
\begin{eqnarray}
\Utrans{\left( \ahat_k + \adag_k \right)^2} \cos 2\Theta_k(t)
=&& \frac{1}{2} \left( \ahat_k^2 + \adagtwo_k \right)
\nonumber \\
&&- g'_+ 
\left( e^{i\Thetakpt} \ahat_k \aphat + \hc \right)
- s_k g'_- 
\left( e^{i\Thetakmt} \ahat_k \amhat + \hc \right)
\nonumber \\
&&+ (\text{other oscillating terms})
+ O(g'^2_\pm), 
\end{eqnarray}
\begin{eqnarray}
\Utrans{\apdag \aphat} 
= \apdag \aphat
+ g'_+ \left( \sumkfour e^{-i\Thetakpt} \apdag \ahat_k + \hc \right)
+ O(g'^2_+), 
\end{eqnarray}
\begin{eqnarray}
\Utrans{\aphat^2} 
= (\text{other oscillating terms}), 
\end{eqnarray}
\begin{eqnarray}
\Utrans{\amdag \amhat} 
= \amdag \amhat
+ g'_- \left( \sumkfour s_k e^{-i\Thetakmt} \amdag \ahat_k + \hc \right)
+ O(g'^2_-), 
\end{eqnarray}
\begin{eqnarray}
\Utrans{\left(\amhat + \amdag \right)^4} 
=&& 6 \amdagtwo \amhat^2 - 12\amdag \amhat + 3
+12g'_- 
\left[ 
\sumkfour s_k e^{-i\Thetakmt} \amdag \left( \amdag \amhat + 1 \right)\ahat_k + \hc 
\right]
\nonumber \\
&&+ (\text{other oscillating terms}) + O(g'^2_-), 
\end{eqnarray}
\begin{eqnarray}
\Utrans{\left( \adag_k - \ahat_k \right) \left( \aphat - \apdag \right) } 
=&& e^{i\Thetakpt} \adag_k \aphat + \hc + (\text{other oscillating terms})
+ O(g'_\pm), 
\label{eq:unitary_trans_adagkaphat}
\end{eqnarray}
and
\begin{eqnarray}
\Utrans{\left(\adag_k - \ahat_k\right) \left(\amhat - \amdag\right)} 
=&& e^{i\Thetakmt} \adag_k \amhat + \hc + (\text{other oscillating terms})
+O(g'_\pm).
\label{eq:unitary_trans_adagkamhat}
\end{eqnarray}
In Eqs.~(\ref{eq:unitary_trans_adagkaphat}) and (\ref{eq:unitary_trans_adagkamhat}), we have not explicitly written the terms of the first order of $g'_\pm$. We do not have to know them to transform $\Hhat$ up to the first order of $g'_\pm$, because $( \adag_k - \ahat_k ) ( \aphat - \apdag )$ and $( \adag_k - \ahat_k ) ( \aphat - \apdag )$ in $\Hhat$ have coefficients $g_+$ and $g_-$, respectively. Equations~(\ref{eq:unitary_trans_adagkaphat}) and (\ref{eq:unitary_trans_adagkamhat}) show that the terms in $\Hhat_g$ obtain $e^{i \Thetakpmt}$ via the transformation as mentioned above. Other oscillating terms in these equations originate from the other coupling terms, e.g., $\adag_k \apdag$. $\Hhat$ is then transformed as
\begin{eqnarray}
&&\frac{1}{\hbar}\Utrans{\Hhat}
\nonumber \\
=&& \sumkfour \left[
\left( \omega + K \right) \adag_k \ahat_k 
-\frac{K}{2} \adagtwo_k \ahat^2_k
+ \frac{p}{2} \left( \adagtwo_k + \ahat^2_k \right)
\right]
+ \omega_+ \apdag \aphat 
+ \left( \omega_- + \Km \right) \amdag \amhat 
- \frac{\Km}{2} \amdagtwo \amhat^2
\nonumber \\
&&+ \sumkfour e^{i\Thetakpt} \Big\{
g_+ \adag_k \aphat 
- g'_+  \left[ 
\omega \adag_k \aphat 
- K \adag_k\left( \adag_k \ahat_k + 1 \right) \aphat 
+ p \ahat_k \aphat 
- \omega_+ \adag_k \aphat 
\right]
\Big\} 
+ \hc
\nonumber \\
&&+ \sumkfour s_k \eiThetakt \Big\{
g_- \adag_k \amhat 
\nonumber \\
&&- g'_-  \left[ 
\omega \adag_k \amhat 
- K \adag_k\left( \adag_k \ahat_k + 1 \right) \amhat 
+ p \ahat_k \amhat 
- \omega_- \adag_k \amhat
+ \Km \adag_k \left( \amdag \amhat + 1 \right) \amhat
\right]
\Big\}
+ \hc
\nonumber \\
&&+ (\text{other oscillating terms}) +O(g_\pm g'_\pm) + O(g'^2_\pm)
\nonumber \\
=&& \sumkfour \left[
\left( \omega + K \right) \adag_k \ahat_k 
-\frac{K}{2} \adagtwo_k \ahat^2_k
+ \frac{p}{2} \left( \adagtwo_k + \ahat^2_k \right)
\right]
+ \omega_+ \apdag \aphat 
+ \left( \omega_- + \Km \right) \amdag \amhat 
- \frac{\Km}{2} \amdagtwo \amhat^2
\nonumber \\
&&+ \sumkfour e^{i\Thetakpt} 
\left\{
\left[ g_+ - g'_+ \left( \omega - \omega_+ - K \right) \right] \adag_k
+ g'_+ \left( K \adagtwo_k - p \right) \ahat_k
\right\}
\aphat
+ \hc
\nonumber \\
&&+ \sumkfour s_k e^{i\Thetakmt} \left( 
\left\{
\left[ g_- - g'_-  \left( \omega - \omega_- - K + \Km  \right) \right] \adag_k 
+ g'_- \left( K \adagtwo_k - p \right) \ahat_k 
\right\}
\amhat
- g'_- \Km \adag_k \amdag \amhat^2
\right)
+ \hc
\nonumber \\
&&+ (\text{other oscillating terms}) +O(g_\pm g'_\pm) + O(g'^2_\pm).
\label{eq:UrdaggerUgdaggerHhatUrUg}
\end{eqnarray}

We tune $g'_\pm$ so that the terms with $e^{i\Thetakpmt}$ for the first order of $g_\pm$ or $g'_\pm$ in Eq.~(\ref{eq:UrdaggerUgdaggerHhatUrUg}) are cancelled out, where the ones with $g_\pm$ are from $\Hhat_g$ [Eq.~(\ref{eq:Hhatg})]. Note that the coupling terms in $\Hhat_g$ also derive higher order ones of $g_\pm$ and $g'_\pm$ with $e^{i\Thetakpmt}$. In particular, we can find them in the terms for the third order. We do not consider such higher order terms when determining $g'_\pm$. Evaluating the correction due to the higher order ones is left as future work. A part of the above terms for the first order of $g_\pm$ and $g'_\pm$ is cancelled out by setting
\begin{eqnarray}
g'_+
&&= \frac{g_+}{\omega - \omega_+ - K}, 
\label{eq:g'+_app}
\\
g'_-
&&= \frac{g_-}{\omega - \omega_- - K + \Km }, 
\label{eq:g'-_app}
\end{eqnarray}
which correspond to Eqs.~(\ref{eq:g'+}) and (\ref{eq:g'-}), respectively. 
The other three terms remain (i.e., $g'_+(K\adagtwo_k - p)\ahat_k \aphat$, $g'_-(K\adagtwo_k - p)\ahat_k \amhat$, $g'_-\Km \adag_k \amdag \amhat^2$), 
but they can be ignored for the ground and low-lying excited states for typical setups. 
When the detuning $\omega - \omega_{p,k}/2$ is smaller than $p/K$, 
the ground state of JPO $k$ without coupling to the coupler in the frame rotating at $\omega_{p,k}/2$ is well described by the superposition of coherent states $\ket{\pm\alpha_0}$, 
where $\alpha_0 = \sqrt{p/K}$~\cite{H.Goto2016Feb, Z.Wang2019}. 
The coherent state is defined by $\ahat \ket{\alpha} = \alpha \ket{\alpha}$~\cite{D.Walls2008}. For these states, the first two remaining terms can be ignored, 
because $\bra{\pm\alpha_0}(K\adagtwo_k - p)\ahat_k \ket{\pm\alpha_0} = \bra{\pm\alpha_0}(K\alpha_0^2 - p)\ahat_k \ket{\pm\alpha_0} = 0$. 
The coupler distorts the superposition of $\ket{\pm\alpha_0}$, but its contribution to the above two terms, which have coefficients $g'_\pm$, becomes higher order. 
It is also reasonable to restrict ourselves to investigate the modes of the coupler lying on the two lowest energy states, i.e., the vacuum or one-photon state, where $g'_-\Km \adag_k \amdag \amhat^2$ vanishes. 
Therefore, we can cancel out the coupling of JPOs and the coupler up to the first order of $g'_\pm$ or $g_\pm$ by using $g'_\pm$ in Eqs.~(\ref{eq:g'+_app}) and (\ref{eq:g'-_app}). 
The transformation with the above $g'_\pm$ will derive effective higher order couplings between JPOs and the coupler.

Let us calculate the unitary transformation of every term in $\Hhat$ [Eq.~(\ref{eq:Hhat})] again, but now we use $g'_\pm$ given by Eqs.~(\ref{eq:g'+_app}) and (\ref{eq:g'-_app}) and focus only on terms without oscillating phases. Oscillating terms will be ignored under the rotating-wave approximation. We need to keep in mind the condition of pump frequencies given by Eq.~(\ref{eq:pump_frequency}) when calculating $\Utrans{(\amhat + \amdag )^4}$. We then have
\begin{eqnarray}
\Utrans{\adag_k \ahat_k}
=&& 
\left(1 - g'^2_+ - g'^2_- + \frac{7}{12} g'^4_+ + \frac{7}{12} g'^4_- + \frac{1}{2}g'^2_+ g'^2_- \right) \adag_k \ahat_k 
+ \sum_{l \neq k} \left( \frac{g'^4_+}{4} + \frac{g'^4_-}{4} + \frac{1}{2} g'^2_+ g'^2_- s_k s_l \right) \adag_l \ahat_l
\nonumber \\
&&+ \left( g'^2_+ - \frac{4}{3} g'^4_+ \right) \apdag \aphat
+ \left( g'^2_- - \frac{4}{3} g'^4_- \right) \amdag \amhat
+ (\text{oscillating terms})
+ O(g'^5_\pm), 
\end{eqnarray}
\begin{eqnarray}
&&\Utrans{\left( \ahat_k + \adag_k \right)^4} 
\nonumber \\
=&& \left( 1 - 2 g'^2_+ -2 g'^2_- + \frac{13}{6} g'^4_+ + \frac{13}{6} g'^4_- + 3 g'^2_+ g'^2_- \right) 
\left( 6 \adagtwo_k \ahat_k^2 - 12\adag_k \ahat_k \right)
\nonumber \\
&&+ g'^4_+ \left( 6 \apdagtwo \aphat^2 - 12\apdag \aphat \right)
+ g'^4_- \left( 6 \amdagtwo \amhat^2 - 12\amdag \amhat \right)
\nonumber \\
&& + 6 \left( 2\adag_k \ahat_k + 1 \right) 
\left[
\sum_{l\neq k} \left( \frac{g'^4_+}{4} + \frac{g'^4_-}{4} + \frac{1}{2} g'^2_+ g'^2_- s_k s_l \right) \left( 2 \adag_l \ahat_l + 1 \right) 
+ g'^2_+ \left( 2 \apdag \aphat + 1 \right) 
+ g'^2_- \left( 2 \amdag \amhat + 1 \right) 
\right]
\nonumber \\
&&+ 6 g'^2_+ g'^2_- 
\left( 2 \apdag \aphat + 1 \right)\left( 2 \amdag \amhat + 1 \right)
+ (\text{oscillating terms}) + O(g'^5_\pm)
\nonumber \\
=&& - 12 \left( 1 - 3 g'^2_+ -3 g'^2_- + \frac{17}{12} g'^4_+ + \frac{17}{12} g'^4_- + \frac{1}{2} g'^2_+ g'^2_- \right) 
\adag_k \ahat_k
+ 6 \left( 1 - 2 g'^2_+ -2 g'^2_- + \frac{13}{6} g'^4_+ + \frac{13}{6} g'^4_- + 3 g'^2_+ g'^2_- \right) 
\adagtwo_k \ahat_k^2 
\nonumber \\
&&+ 12 \sum_{l\neq k} \left( \frac{g'^4_+}{4} + \frac{g'^4_-}{4} + \frac{1}{2} g'^2_+ g'^2_- s_k s_l \right) \adag_l \ahat_l 
\nonumber \\
&&+ 12 \left( g'^2_+ - g'^4_+ + g'^2_+ g'^2_- \right) \apdag \aphat
+ 6 g'^4_+ \apdagtwo \aphat^2 
+ 12 \left( g'^2_- - g'^4_- + g'^2_+ g'^2_- \right) \amdag \amhat
+ 6 g'^4_- \amdagtwo \amhat^2 
\nonumber \\
&& + 12 \left[
\sum_{l\neq k} \left( \frac{g'^4_+}{2} + \frac{g'^4_-}{2} + g'^2_+ g'^2_- s_k s_l \right) \adag_l \ahat_l 
+ 2g'^2_+ \apdag \aphat + 2g'^2_- \amdag \amhat  
\right] \adag_k \ahat_k 
+ 24 g'^2_+ g'^2_- \apdag \aphat \amdag \amhat 
\nonumber \\
&& + (\text{oscillating terms}) + O(g'^5_\pm), 
\end{eqnarray}
\begin{eqnarray}
\Utrans{\left( \ahat_k + \adag_k \right)^2} \cos 2\Theta_k(t)
=&& \frac{1}{2} 
\left( 1 - g'^2_+ - g'^2_- + \frac{7}{12} g'^4_+ + \frac{7}{12} g'^4_- + \frac{1}{2}g'^2_+ g'^2_- \right)
\left( \ahat_k^2 + \adagtwo_k \right)
\nonumber \\
&&+ (\text{oscillating terms})
+ O(g'^5_\pm), 
\end{eqnarray}
\begin{eqnarray}
\Utrans{\apdag \aphat} 
= \left( 1 - 4 g'^2_+ + \frac{16}{3} g'^4_+ \right) \apdag \aphat
+ \left( g'^2_+ - \frac{4}{3} g'^4_+ \right) \sumkfour \adag_k \ahat_k
+ (\text{oscillating terms})
+ O(g'^5_+), 
\end{eqnarray}
\begin{eqnarray}
\Utrans{\aphat^2} 
= (\text{oscillating terms}), 
\end{eqnarray}
\begin{eqnarray}
\Utrans{\amdag \amhat} 
= \left( 1 - 4 g'^2_- + \frac{16}{3} g'^4_- \right) \amdag \amhat
+ \left( g'^2_- - \frac{4}{3} g'^4_- \right) \sumkfour \adag_k \ahat_k
+ (\text{oscillating terms})
+ O(g'^5_-), 
\end{eqnarray}
\begin{eqnarray}
&&\Utrans{\left(\amhat + \amdag \right)^4}
\nonumber \\ 
=&& \left( 1 - 8 g'^2_- + \frac{80}{3} g'^4_- \right) 
\left( 6 \amdagtwo \amhat^2 - 12\amdag \amhat \right)
+ g'^4_- \sumkfour \left( 6 \adagtwo_k \ahattwo_k - 12\adag_k \ahat_k \right)
\nonumber \\
&& + \left( 6 g'^2_- - 32 g'^4_- \right)
\left( 2 \amdag \amhat + 1 \right) \sumkfour \left( 2 \adag_k \ahat_k + 1 \right)
+ 6 g'^4_- 
\sum_{k<l} \left( 2 \adag_k \ahat_k + 1 \right) \left( 2 \adag_l \ahat_l + 1 \right)
\nonumber \\
&&+ 24 g'^4_-
\left( e^{i \sumkfour s_k \theta_{p, k}/2} \adag_1 \adag_2 \ahat_3 \ahat_4 
+ \text{h.c.} 
\right)
+ (\text{oscillating terms}) + O(g'^5_-)
\nonumber \\
=&& - 12 \left( 1 - 12 g'^2_- + 48 g'^4_- \right) \amdag \amhat
+6 \left( 1 - 8 g'^2_- + \frac{80}{3} g'^4_- \right) \amdagtwo \amhat^2 
+ 12 \left( g'^2_- - \frac{10}{3} g'^4_- \right) \sumkfour \adag_k \ahat_k
+ 6 g'^4_- \sumkfour \adagtwo_k \ahattwo_k 
\nonumber \\
&&+ 12\left( 2g'^2_- - \frac{32}{3} g'^4_- \right)
\amdag \amhat \sumkfour \adag_k \ahat_k
 + 24 g'^4_- 
\sum_{k<l} \adag_k \ahat_k \adag_l \ahat_l
\nonumber \\
&&+ 24 g'^4_-
\left( e^{i \sumkfour s_k \theta_{p, k}/2} \adag_1 \adag_2 \ahat_3 \ahat_4 
+ \text{h.c.} 
\right)
+ (\text{oscillating terms}) + O(g'^5_-), 
\end{eqnarray}
\begin{eqnarray}
\Utrans{\left( \adag_k - \ahat_k \right) \left( \aphat - \apdag \right) } 
=&& \left( 2g'_+ - \frac{4}{3} g'^3_+ \right) \adag_k \ahat_k
- \sumlfour \left( g'^3_+ + g'_+ g'^2_- s_k s_l \right) \adag_l \ahat_l 
- \left( 2g'_+ - \frac{16}{3} g'^3_+ \right) \apdag \aphat
\nonumber \\
&&+ (\text{oscillating terms}) + O(g'^5_\pm), 
\end{eqnarray}
and
\begin{eqnarray}
\Utrans{\left(\adag_k - \ahat_k\right) \left(\amhat - \amdag\right)} 
=&& s_k \left( 2g'_- - \frac{4}{3} g'^3_- \right) \adag_k \ahat_k
- s_k \sumlfour \left( g'^3_- + g'^2_+ g'_- s_l s_k \right) \adag_l \ahat_l 
- s_k \left( 2g'_- - \frac{16}{3} g'^3_- \right) \amdag \amhat
\nonumber \\
&&+ (\text{oscillating terms}) +O(g'^5_\pm), 
\end{eqnarray}
where constants have been dropped.
These terms with coefficients in $\Hhat$ [Eq.~(\ref{eq:Hhat})] 
and the terms in Eq.~(\ref{eq:UrdaggerUrdot}) 
constitute 
\begin{eqnarray}
\frac{\Hhat'}{\hbar}
=&& \frac{1}{\hbar} \hat{U}_r^\dagger \hat{U}_g^\dagger \Hhat \hat{U}_g \hat{U}_r - i \hat{U}_r^\dagger \dot{\hat{U}}_r
\nonumber \\
=&&
\sumkfour 
\left[
\Delta_k \adag_k \ahat_k 
- \frac{K'}{2} \adagtwo_k \ahattwo_k 
+ \frac{p}{2} \left( \adagtwo_k + \ahattwo_k \right)
\right]
+\Delta_+ \apdag \aphat 
- \frac{\Kp'}{2} \apdagtwo \aphattwo
+\Delta_- \amdag \amhat 
- \frac{\Km'}{2} \amdagtwo \amhattwo
\nonumber
\\
&&- \gfour \left( 
e^{i \sumkfour s_k \theta_{p,k} / 2} \adag_1 \adag_2 \ahat_3 \ahat_4 
+ e^{-i \sumkfour s_k \theta_{p,k} / 2} \ahat_1 \ahat_2 \adag_3 \adag_4 
\right)
\nonumber \\
&&- \sum_{k<l} \gtwo_{kl} \adag_k \ahat_k \adag_l \ahat_l 
- \gtwo_{\text{J}+} \sumkfour \adag_k \ahat_k \apdag \aphat
- \gtwo_{\text{J}-} \sumkfour \adag_k \ahat_k \amdag \amhat
- \gtwo_{+-} \sumkfour \apdag \aphat \amdag \amhat,
\end{eqnarray}
where
\begin{eqnarray}
\Delta_k 
=&& \omega \left(1 - g'^2_+ - g'^2_- + \frac{4}{3} g'^4_+ + \frac{4}{3} g'^4_- \right) 
+ K \left( 1 - 3 g'^2_+ -3 g'^2_- + \frac{2}{3} g'^4_+ + \frac{2}{3} g'^4_- + g'^2_+ g'^2_- \right) 
\nonumber \\
&&+ \omega_+ \left( g'^2_+ - \frac{4}{3} g'^4_+ \right)
+ \omega_- \left( g'^2_- - \frac{4}{3} g'^4_- \right)
+ \Km \left( - g'^2_- + \frac{10}{3} g'^4_- \right) 
\nonumber \\
&&+ g_+ \left( 2g'_+ - \frac{16}{3} g'^3_+ \right) 
+ g_- \left( 2g'_- - \frac{16}{3} g'^3_- \right)
- \frac{\omegap{k}}{2}
\nonumber \\
=&& \omega \left(1 + g'^2_+ + g'^2_- - 4 g'^4_+ - 4 g'^4_- \right) 
+ K \left( 1 - 5 g'^2_+ -5 g'^2_- + 6 g'^4_+ + 6 g'^4_- + g'^2_+ g'^2_- \right) 
\nonumber \\
&&- \omega_+ \left( g'^2_+ - 4 g'^4_+ \right)
- \omega_- \left( g'^2_- - 4 g'^4_- \right)
+ \Km \left( g'^2_- -2 g'^4_- \right) 
- \frac{\omegap{k}}{2}, 
\end{eqnarray}
\begin{eqnarray}
K' 
=&& \left( 1 - 2 g'^2_+ -2 g'^2_- + \frac{13}{6} g'^4_+ + \frac{13}{6} g'^4_- + 3 g'^2_+ g'^2_- \right) K  
+ \frac{g'^4_-}{2} \Km, 
\end{eqnarray}
\begin{eqnarray}
\Delta_+ 
=&& 4 \left( g'^2_+ - \frac{4}{3} g'^4_+ \right) \omega 
- 4 \left( g'^2_+ - g'^4_+ + g'^2_+ g'^2_- \right) K
- 4  \left( g'^2_+ - \frac{4}{3} g'^4_+ \right) \omega_+
- 4 \left( 2g'_+ - \frac{16}{3} g'^3_+ \right) g_+
\nonumber \\
=&& -4 \left( g'^2_+ - 4 g'^4_+ \right) \omega 
+ 4 \left( g'^2_+ - \frac{13}{3} g'^4_+ - g'^2_+ g'^2_- \right) K
+ 4 \left( g'^2_+ - 4 g'^4_+ \right) \omega_+, 
\end{eqnarray}
\begin{eqnarray}
\Delta_- 
=&& 4 \left( g'^2_- - \frac{4}{3} g'^4_- \right) \omega 
- 4 \left( g'^2_- - g'^4_- + g'^2_+ g'^2_- \right) K
- 4  \left( g'^2_- - \frac{4}{3} g'^4_- \right) \omega_- 
+ \left( 1 - 12 g'^2_- + 48 g'^4_- \right) \Km
- 4 \left( 2g'_- - \frac{16}{3} g'^3_- \right) g_-
\nonumber \\
=&& -4 \left( g'^2_- - 4 g'^4_- \right) \omega 
+ 4 \left( g'^2_- - \frac{13}{3} g'^4_- - g'^2_+ g'^2_- \right) K
+ 4 \left( g'^2_- - 4 g'^4_- \right) \omega_- 
+ \left( 1 - 20 g'^2_- + \frac{208}{3} g'^4_- \right) \Km, 
\end{eqnarray}
\begin{eqnarray}
\Kp'
=&& 4 g'^4_+ K, 
\end{eqnarray}
\begin{eqnarray}
\Km'
=&& 4 g'^4_- K 
+  \left( 1 - 8g'^2_- + \frac{80}{3}g'^4_- \right) \Km, 
\end{eqnarray}
\begin{eqnarray}
\gfour
=&& 2 g'^4_- \Km, 
\end{eqnarray}
\begin{eqnarray}
\gtwo_{kl}
=&& \left( 2 g'^4_+ + 2 g'^4_- + 
4 g'^2_+ g'^2_- s_k s_l 
\right) K 
+ 2 g'^4_- \Km
\end{eqnarray}
\begin{eqnarray}
\gtwo_{\text{J}+}
=&& 2 g'^2_+ K,
\end{eqnarray}
\begin{eqnarray}
\gtwo_{\text{J}-}
=&& 2 g'^2_- K 
+ \left( 2 g'^2_- - \frac{8}{3}g'^4_- \right) \Km,
\end{eqnarray}
and
\begin{eqnarray}
\gtwo_{+-}
=&& 2 g'^2_+ g'^2_- K.
\end{eqnarray}
We have ignored constants, oscillating terms, 
and the terms of $O(g'^5_\pm)$.
This Hamiltonian is diagonal
for the two resulting modes of coupler.
Thus, 
if we set the initial state of the coupler to a Fock state, 
it will remain 
even if the parameters are varied over time.
Let us assume the coupler in the vacuum state.
The above Hamiltonian then reduces to 
\begin{eqnarray}
\frac{\Hhat'}{\hbar}
=&& \sumkfour 
\left[
\Delta_k \adag_k \ahat_k 
- \frac{K'}{2} \adagtwo_k \ahattwo_k 
+ \frac{p}{2} \left( \adagtwo_k + \ahattwo_k \right)
\right]
\nonumber
\\
&&- \gfour \left( 
e^{i \sumkfour s_k \theta_{p,k} / 2} \adag_1 \adag_2 \ahat_3 \ahat_4 
+ e^{-i \sumkfour s_k \theta_{p,k} / 2} \ahat_1 \ahat_2 \adag_3 \adag_4 
\right)
- \sum_{k<l} \gtwo_{kl} \adag_k \ahat_k \adag_l \ahat_l.
\end{eqnarray}
This is the effective Hamiltonian shown in the main text [Eq.~(\ref{eq:Hhat'})].

\end{widetext}

\end{document}